\documentclass[sigconf]{acmart}

\usepackage{tabularx}
\usepackage{caption}
\usepackage{subcaption}
\usepackage{array}

\newcolumntype{C}[1]{>{\centering\arraybackslash}p{#1}}

\AtBeginDocument{%
  \providecommand\BibTeX{{%
    \normalfont B\kern-0.5em{\scshape i\kern-0.25em b}\kern-0.8em\TeX}}}

\copyrightyear{2022}
\acmYear{2022}
\setcopyright{acmcopyright}\acmConference[UIST '22]{The 35th Annual ACM Symposium on User Interface Software and Technology}{October 29-November 2, 2022}{Bend, OR, USA}
\acmBooktitle{The 35th Annual ACM Symposium on User Interface Software and Technology (UIST '22), October 29-November 2, 2022, Bend, OR, USA}
\acmPrice{15.00}
\acmDOI{10.1145/3526113.3545621}
\acmISBN{978-1-4503-9320-1/22/10}

\begin{document}

\title{Opal: Multimodal Image Generation for News Illustration}


\author{Vivian Liu}
\email{vivian@cs.columbia.edu}
\affiliation{%
  \institution{Columbia University}
  \city{New York}
  \state{New York}
  \country{USA}
}

\author{Han Qiao}
\email{h.qiao@mail.utoronto.ca}
\affiliation{%
  \institution{University of Toronto}
  \city{Toronto}
  \state{Ontario}
  \country{Canada}
}

\author{Lydia B. Chilton}
\email{chilton@cs.columbia.edu}
\affiliation{%
  \institution{Columbia University}
  \city{New York}
  \state{New York}
  \country{USA}
}
\renewcommand{\shortauthors}{Liu, Qiao, and Chilton}

\begin{abstract}
Advances in multimodal AI have presented people with powerful ways to create images from text. Recent work has shown that text-to-image generations are able to represent a broad range of subjects and artistic styles. However, finding the right visual language for text prompts is difficult. In this paper, we address this challenge with Opal, a system that produces text-to-image generations for news illustration. Given an article, Opal guides users through a structured search for visual concepts and provides a pipeline allowing users to generate illustrations based on an article's tone, keywords, and related artistic styles. Our evaluation shows that Opal efficiently generates diverse sets of news illustrations, visual assets, and concept ideas. Users with Opal generated two times more usable results than users without. We discuss how structured exploration can help users better understand the capabilities of human AI co-creative systems. 

\end{abstract}


\begin{teaserfigure}
  \includegraphics[width=\textwidth]{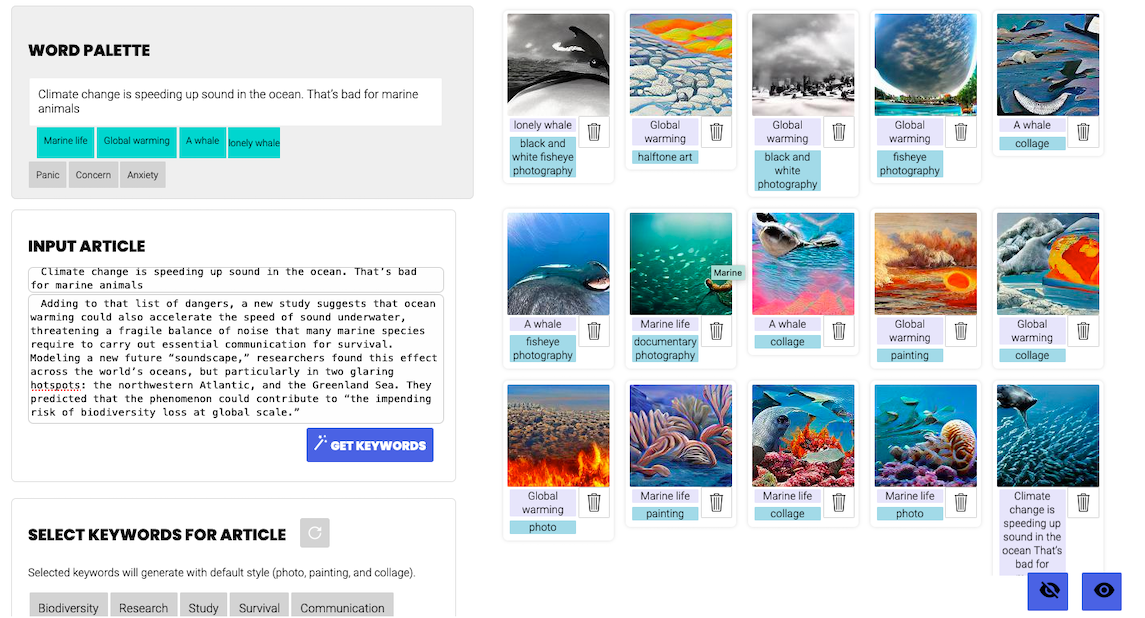}
  \caption{A screenshot of the Opal system, which helps users create news illustrations using a text-to-image generative AI model. The system here has generated a gallery of images for an article on "climate change". The participant is guided through the generation process with a structured pipeline of suggestions based off of GPT-3 generated keywords, tones, and styles. }
  \Description{A screenshot of the Opal system. On the left side, different tools that guide user input and selection are shown including a section called "Input Article" and a section called "Select Keywords for Article." On the right side, a five by three gallery of fifteen generated images are shown. The images are generated for an article about how global warming slows the speed of sound underwater and how that will affect marine lives.}
  \label{fig:teaser}
\end{teaserfigure}

\maketitle

\section{Introduction}
Text-to-image generative frameworks allow users to generate images based on text. These frameworks utilize deep learning models that have been pre-trained on large magnitudes of data, allowing users the freedom to experiment with a near infinite number of visual concepts. However, while we can compose any number of visual concepts in text and language, it is not guaranteed that any prompt passed through a text-to-image AI will produce a quality outcome, and what models are capable of generating is opaque to the user. 

Recent empirical work in \cite{liu2021design} has shown that structured prompts such as "\{SUBJECT\} in the style of \{STYLE\}" can produce consistent results across a wide variety of subjects and styles. Furthermore, styles with very salient visual characteristics (such as a specific artist’s name) can help guide users towards high quality aesthetic outcomes. However, these models still perform variably depending on the style and subject matter. One core usability problem with these frameworks is that understanding what combinations of subjects and styles the model is capable of generating is often a random, trial and error process. When the generations fail, they often result in generations with distorted, uncanny, and unnatural compositions. Because these systems are also stochastic, we explore the challenge of translating text into images in a systematic, structured, and efficient way.

An area that would greatly benefit text-to-image models is news illustration. News illustration refers to the practice of creating visuals that accompany news and journalism feature pieces. News illustrations often capture facets of the article such as the emotion, levity, and subject matter with visual metaphors and conceptual illustrations. News illustrators often work under the breakneck pace of a news cycle and the pressure for news to be presented in a timely, relevant, and eye-catching manner.  They often have to come out fast and be created rapidly after the article is written---or even concurrently.

To help news illustrators efficiently create illustrations, we present Opal, a system that guides users through a structured search for visual concepts beginning with article text. We organize and streamline prompt ideation for text-to-image generation by providing three pipelines that allow users to engage with the article in all its facets: its emotional tone, its subject matter, and its goal illustration style.  At each stage of the pipeline, we employ techniques and leverage advancements from natural language processing such as semantic search and methods of prompt engineering GPT-3 to query it as a knowledge base and association network. We present the following set of contributions:

\begin{itemize}
\item Opal, a system that guides users through a structured search for visual concepts
\item Automatic techniques of structured prompt engineering
 to help translate text into images
 \item Annotation studies demonstrating that suggestions prompted from large language models can retrieve keywords, tones, styles, and associated visual concepts, while approaching human benchmarks with significantly less effort.
 
\item User studies demonstrating the efficacy of this structured prompt engineering within in a system. Users with Opal were two times more efficient at generation and generated two times more usable results than users without. 

\end{itemize}

We end by discussing how large language models can structure exploration with text-to-image AI, how such technologies can augment rather than replace the process of illustrators, and how users can systematically experiment to understand generative AI capabilities.

\section{Related Work}

Generative models have evolved greatly over the past decade in tandem with machine learning. Neural style transfer approaches were some of the early and popular seminal approaches. They applied style attributes from a source image to target images.\cite{gatys2015neural} Building on top of this work in the generative adversarial networks (GAN) space, StyleGAN and StyleGAN2 showed that deep learning models could learn from and stochastically vary style attributes within classes of images such as faces. \cite{karras2020analyzing} SPADE was another technique that allowed stylistic variation over high quality images, while keeping the semantic information within those images intact. Within a demonstration of SPADE, users could control style in a continuous dimension–for example, turning a cool cloudy blue ocean horizon into a fiery red sunrise at sea. \cite{park2019SPADE} These works affirm that style has long been a popular method of steering and interacting with generative AI tools. One rationale for why stylization is so compelling as a core functionality for generative systems could be that subject and style have been empirically found within cognitive science literature to be separately processed modes of information. \cite{Kawabata2004NeuralCO,cupchik,augustin}

While style is one vector of interest, a north star goal for human AI co-creation systems has always been declarative generation, the ability to semantically control generative outputs. This goal has been approached in a number of domains from images to music to 3D shapes. \cite{aggarwal2020neurosymbolic, cococo,Chaudhuri:2013:ACC, liu_chilton} For example, within the domain of music, Cococo gave users the ability to control at a local level how happy, sad, or surprising a measure of music was---allowing users to declaratively edit music based on emotion with their concept of “semantic sliders”. 

Language, however, presents one of the highest forms of control as it is understandable by nature. Studies have looked at how language models can be embedded into creativity support systems. \cite{calderwood_how_2018} studied how novelists could interact with language models to write fiction and explored the trade-offs between autocomplete and user control. Both \cite{chakrabarty_mermaid_2021} and \cite{gero_metaphoria_2019} investigated how computational models could connect abstract concepts and symbols to implement metaphor generation. Sparks \cite{sparks} studied how scientists could integrate suggestions from GPT2 in a technical writing process, finding patterns of interaction in the way writers translated AI-generated technical suggestions into short-form science communication.

\subsection{Prompt Engineering in NLP}
The rise of language transformer models \cite{gpt3} and their success in few-shot learning has led to a new form of interaction with AI models known as prompt engineering. Prompt engineering refers to the formal search for prompts that can condition the model to produce more relevant and high quality outcomes. \cite{gwern_2020, reynolds2021prompt} At present, it is currently unknown how prompt engineering compares to fine-tuning and which is better. A similar open research question is whether or not automatically discovered continuous prompts \cite{gao_making_2021} or discrete prompts are better for prompt engineering. \cite{li_prefix-tuning_2021} While automatically tuned prompts have shown modestly better performance on certain benchmarks \cite{li_prefix-tuning_2021}, discrete manually crafted prompts are advantageous in that they are human readable and can draw from existing traditions of work. For example, in the system Sparks \cite{sparks}, the prompt engineering that helped extract ideas from GPT-2 for their system was crafted based on narrative and expository theory. This allowed the prompts to additionally be integrated into the system as structured guidance and ideation for their writing support tool. Going beyond single prompts, \cite{aichains} and Promptchainer \cite{promptchainer} demonstrated how complex tasks can be decomposed and supported by workflows of chained prompts. \cite{promptbasedellen} was one of the first to show the utility of prompts embedded within systems by helping industry professionals conduct prompt-based prototyping of machine learning.

Aside from the GPT family, BERT has also been a longstanding model for the AI community to study and utilize. \cite{devlin2019bert} BERT’s masked language modeling objective has often been employed in the past for model completion---in essence creating something to the effect of a templated prompt. Within \cite{ge2021visual}, BERT was utilized as a part of the text-to-image pipeline. By having BERT produce shape analogies by responding to a masked template, they demonstrated how the commonsense and world knowledge within large pretrained models could be utilized within systems. Likewise in our system, we use GPT-3 by querying it as a knowledge base for keywords, tones, icons, and style knowledge. However, we approach GPT-3 from a prompt engineering perspective and provide a number of prompting solutions that connect text and image together on dimensions beyond just shape.

\subsection{Multimodality}
Our system adds to a long tradition of work within multimodal authoring tools. \cite{sceneseer} and Wordseye \cite{wordseye} were systems that connected language to 3D space, allowing users to use text prompts to generate 3D scenes. \cite{Chaudhuri:2013:ACC} allowed users to define 3D shapes based on adjectives and emotive cues, parameterizing creature shapes by qualities like “cute” or “dangerous”. As tools moved away from statistical-based techniques to GANs, a number of text-to-image architectures \cite{Attngan, mirrorgan, leicagan, tedigan} were proposed. However, many of these architectures were class-conditional, meaning they were constrained to the single-class datasets they were trained upon.

Recently, multimodal systems have found renewed momentum from machine learning advancements in multimodal representation learning. CLIP, one of the latest advancements, demonstrated that images and text pairs can be contrastively learned and optimized such that the image and text embeddings can share a multimodal embedding space. \cite{radford2021learning} Many of the earliest open-source text-to-image generative frameworks to gain traction used CLIP in a discriminator-like fashion, using it in conjunction with a host of generative models from BigGan to VQGAN. \cite{adverb, crowson,vqgan, crowson2022vqgan} Newer methods such as diffusion models have also increased output quality. \cite{nerdyroden, murdock2,crowsondiffusion, dalle} DALL·E 2 \cite{unclip} showed by conditioning on a CLIP text prior, generating CLIP image representations, and decoding those representations either autoregressively or with diffusion, DALL·E 2 could create a text-to-image generator with novel functionalities like image remixing, inpainting, and interpolation. Other state-of-the-art text-to-image architectures have come out in rapid succession.  Make-a-Scene offers users control in the form of semantic layout sketches.\cite{meta_ai} Imagen \cite{imagen} and PARTI \cite{parti} both provide state of the art text-to-image generation capabilities while also introducing comprehensive benchmarks to evaluate architectures on key aspects of visual language such as compositionality, cardinality, spatial relations, etc. DALL-Eval found that text-to-image generators still manifest social and gender biases. \cite{dall-eval}

To our knowledge, no text-to-image system with a focus on a specific design task like news illustration exists yet. The closest work in terms of prior systems to ours include: 1) Chatpainter \cite{chatpainter}, which generated images through dialogue with a chatbot, 2) Keep Drawing It \cite{keepdrawingit}, which generated images based on iterative linguistic feedback, and 3) Stolen Elephant, a multimodal authoring tool that leveraged text, image, and aural prompts to help inspire creative writing  \cite{stolenelephant}.

\section{Formative Study}

In order to understand how a text-to-image model could best augment a news illustrator's process, we conducted a co-design process with three news illustrators on a weekly basis over the course of two months. We recruited these illustrators by connecting with a local newspaper that all three worked for. These news illustrators had experiences spanning both staff illustrator and illustration editor positions. All had extensive backgrounds in art (over five years of art practice).

Week to week, we had illustrators generate sets of text-to-image generations using VQGAN+CLIP \cite{crowson2022vqgan} for chosen news articles. These generations explored prompts structured in the template "\{SUBJECT\} in the style of \{STYLE\}". By parameterizing the prompts in this way, we systematically searched for qualities of the successful prompts that represented the articles well. These text-to-image generations were then discussed as a group for their holistic aesthetic qualities and whether or not they were able to express the intended subject and styles of the prompts. During this process, we asked the illustrators to compare and contrast this AI approach with their usual process.
 
\subsection{Findings}

\subsubsection{Traditional News Illustration Process} We first learned how news illustrators traditionally crafted illustrations. All the illustrators confirmed that they are generally given partial information from work-in-progress article drafts to start with: chunks of article text, keywords, working article titles, and high-level description. More often than not, news illustrators do not get well-formed, complete articles to work off of. One of the news illustrators said that they often received text from early paragraphs or later paragraphs of their assigned articles. They rarely received titles, because those generally tended to be come up with last.

\subsubsection{Understanding How News Illustrations Combine Subjects and Styles}

Early on, we found that text-to-image models were not nuanced enough to simply take in a news headline and output a high-quality cover image. Text prompts had to contain more visual language, which was why part of our process was to let illustrators explore prompts grounded in concrete, highly visual subject and style information.

\begin{figure}
    \centering
    \includegraphics[width=0.5\textwidth]{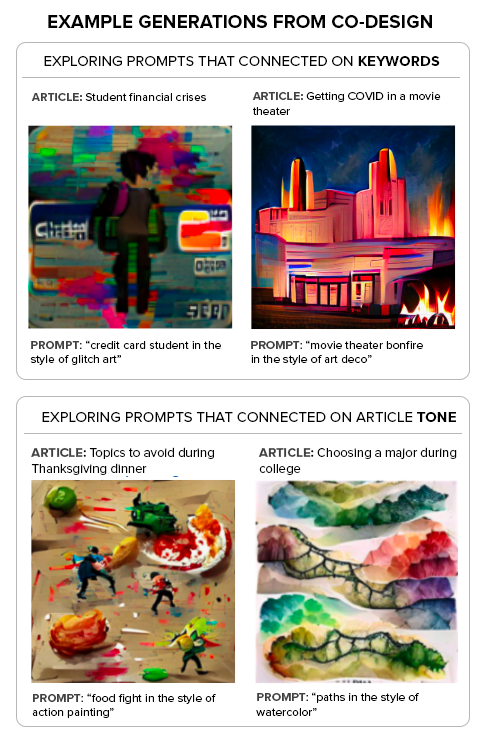}
    \caption{Text-to-image generations that were successful with news illustrators during the co-design process. These generations captured design patterns discussed in the formative study, where subjects and styles were suggested based on keywords and tones. For example, in the top left, "glitch art" was a relevant style chosen by a news illustrator to visually express an article involving credit cards and crisis. In the bottom left, a news illustrator chose an "action painting" style to express chaotic energy, an example of a news illustrator working off of an article tone.}
    \Description{Four images generated from text-to-image model are presented in a 2 by 2 grid. The top left image is generated with the text prompt "Credit card student in the style of glitch art." The image has a person icon in the foreground and a credit card image on the background. The top right image is generated with the text prompt "Movie theatre bonfire in the style of art deco." The image shows a golden movie theatre building with fire on the right side and the roof of the building under a dark blue sky. Both of the top two images are examples of choosing art styles relevant to keywords. The bottom left image is generated with the text prompt "Food fight in the style of action painting." The image shows human figures, pieces of food and color blocks across the whole canvas. The bottom right image is generated using the text prompt "Paths in the style of water color." The image shows light green paths being surrounded by other light color blocks giving it an airy and serene aesthetics. Both of the bottom images are generations where art styles are picked to be relevant to the tones of the subjects.}
    \label{fig:designpatterns}
\end{figure}

We learned that illustrators were approaching style differently when they used text-to-image AI. Traditionally, illustrators tended to stick with their own style by virtue of their expertise. However, illustrators explored more broadly across styles when they produced text-to-image generations. As a group, we gradually found that certain styles worked better than others for news. These styles tended to connect with the subject on two dimensions: on concrete subject matter like keywords and global characteristics like author tone. For example, "glitch art" was a style of art that worked well for articles with subjects related to computers, Internet, and the digital age---connecting on keywords. On the other hand, we found that certain artistic styles were more effective at conveying an emotion more globally across the composition---connecting on tone. For example, "action painting" as an art style created a sense of dynamic movement, positive energy, and excitement that paired well with news articles describing family fights during Thanksgiving dinners, which we can see in the generation of "food fight in the style of action painting" in Figure \ref{fig:designpatterns}. On the other hand, other styles such as Impressionism captured abstract qualities such as happiness, wonder, and tranquility. However, we also learned that prompting text-to-image models with words that were too abstract made it difficult for the model to find purchase in an recognizable subject. To remedy this, we often looked for symbols to represent abstract concepts; for example, we expanded "happiness" to "sun", "beachballs", and "emojis". We used this approach particularly for tone words that tended to be more abstract.

Illustrators liked to attempt styles that often accompanied news articles such as "vector art". However, generating in the "vector art" style tended to produce generations that were hit or miss. Sometimes, they successfully produced thick cartoonish, flat illustrations, but often, they produced blank images when the AI failed to optimize towards the text prompt. Based off of experimentation with the illustrators, we curated a list of styles commonly seen in journalism and pared it down for ones that VQGAN+CLIP could replicate. These styles included cartoon, vector art, street photography, pencil drawing, flat illustration, and so on. We additionally curated a list of styles that happened to perform well in terms of aesthetic quality from VQGAN+CLIP, which built off of previous research that analyzed the stylistic knowledge spanned by VQGAN+CLIP \cite{liu2021design}. In total, we created a list of 125 styles.

\subsection{Generation Usability for News Illustrations}

Throughout the process of analyzing generations, we searched for design patterns that could structure the process of creating high-quality generations. We realized that often times the generations did not necessarily  need to be taken as is. Sometimes they were fantastic standalone, but often their artifacts and flaws could be easily edited away. Generations could be collaged, drawn over, and post-processed in a number of ways as a visual asset. Illustrators commented that they could work around unclear subjects by using generations as backgrounds and editing in subjects from real images, or taking the compositions of generations as ideas. The learnings from these co-design sessions helped us learn about news illustration and justify design choices that were later implemented in Opal.




\subsection{System Design}

\begin{figure*}
    \centering
    \includegraphics[width=\textwidth]{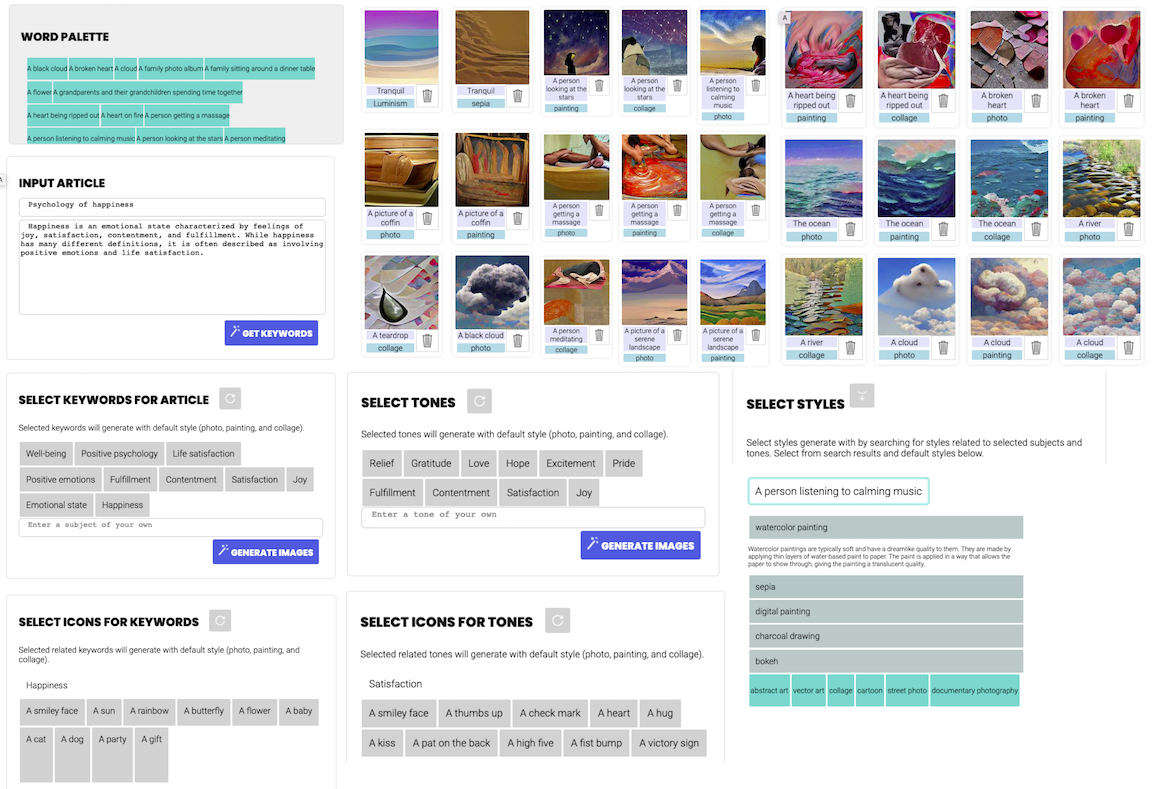}
    \caption{System design: The system takes in headlines and article content. From the news-related text, Opal generates keywords and tones, as well as icons that can visualize those keywords and tones. Opal also allows users to get recommendations for artistic styles to generate in. While this figure shows an exploded view of all the features, the actual system organizes all of the exploration features and pipelines on the left side and the gallery view on the right side.}
    \Description{Multiple screenshots of the system, demonstrating all features in the system, including "Word Palette", "Input Article","Select keywords for article", "Select icons for keywords", "Select tones","Select icons for tones", "Select styles" and on the top right, a grid of 9 by 3 generations is shown. }
    \label{fig:systemdesign}
\end{figure*}

 In our formative co-design study, we learned that news illustrators received text as their prompts. They would first conceptually explore the text and find related phrases and concepts to visually capture the article. Building off of this, we focused on text prompts as the beginning point of exploration and developed simple prompt engineering methods to retrieve keyword, tone, icon, and style suggestions.

\subsubsection{Motivating our choice of GPT-3}
We chose GPT-3 compared to other NLP models because it has been proven to do a variety of NLP tasks (summarization, keywords, ideation, translation, synonyms) with no explicit task-specific training. GPT-3 can perform all those abilities at once, making it a versatile tool to help people with real world tasks. Having seen an Internet-scale amount of data and developed capabilities for zero-shot learning, it can find related words easily and with good accuracy. We used the 'Davinci' model of GPT-3, with 256 tokens returned and 3 for the "best of" hyperparameter; Davinci is the largest and most capable model. 

\subsubsection{Keyword and Tone Suggestion.}
To generate keyword suggestions for some \{ARTICLE TEXT\}, we prompted GPT-3 with the following instruction prompt, “Here are ten keywords for: \{ARTICLE TEXT\}", and parsed the output. 
Likewise, we extracted tones from the text using the following prompt for GPT-3: "Here are ten emotions for: \{ARTICLE TEXT\}”.

\subsubsection{Icon Suggestions for Keywords and Tones.}

We prompted GPT-3 using a prompt template: “Here are 10 icons related to: \{KEYWORD / TONE \}” We chose to look for “icons”, as they exist as abstractions in between the word and image space. This prompt allowed us to utilize GPT-3 as a knowledge base to provide associative knowledge. From these expansions, we were able to collect diverse sets of subjects from an unbounded number of options that would have been otherwise unavailable if we used dataset approaches such as Small World of Words.

\subsubsection{Style Suggestion}
Given a user query, we recommended styles based on how well that query matched sentences within a dataset of style information. This dataset was collected from GPT-3; we scraped GPT-3 as if it were a knowledge base for artistic styles using the following prompt: “What are some of the defining characteristics of \{STYLE\} as an art style?” We conducted this scrape using the "best-of" parameter set at 3 and with the maximum number of tokens set to 256. For example, for "gothic art", GPT-3 returned that "some common characteristics of Gothic art are intricate designs, often featuring pointed arches; tall, thin spires; and large stained glass windows. Gothic art is often associated with the Gothic architecture, which is characterized by its pointed arches, ribbed vaults, and flying buttresses." We conducted this scrape for all 125 styles from the styles list developed during co-design.

These sentences were saved as results to search over using Sentence-BERT's \cite{sbert} approach for asymmetric semantic search. For example, upon inputting a query such as “dark and moody”, users were presented with results such as “gothic art”, “baroque”, “black and white photography”, “photo negative” based on semantic similarity of "dark and moody" to sentences of these styles that GPT-3 had scraped. 

\subsubsection{Text-to-Image Generation.}

For text-to-image generation, we used the checkpoint and configuration of VQGAN+CLIP that was pretrained on Imagenet with a codebook size of 16384 \cite{nerdyroden}. Each prompted image was generated to be 256x256 pixels and iterated on for 100 steps on either a local NVIDIA GeForce RTX 3080 GPU or a remote Tesla V100 GPU.

\subsubsection{User Interface.}

Opal, seen in Figure \ref{fig:systemdesign} is an interface composed of two components: Oeuvre and Palette. An oeuvre is by definition "the works of a painter, composer, or author regarded collectively". The oeuvre provides users with a birds-eye view of all the generations they have created, and the generations stream in in real time as they are generated. The Palette provides users with a pipeline of tools to help create a news illustration. The user begins with the Article Area, which asks the user to input article text, which could be short one-line descriptions or long-form text. 

Upon submitting article text, ten keywords would populate the Keyword Area, summarizing the article. Ten tones would likewise populate the Tone Area. Keywords and tones could be expanded one by one for visual icons relevant to the selected word. They were both retrieved by the icon suggestion prompt method. Users could generate by selecting keywords, tones, and icons they wanted to try and hitting 'Generate'. These selections were automatically generated with three variations of the prompts, to show users variety. These variations spanned different default styles: "\{SELECTED WORD\} in the style of a \{photo / collage / painting \}". We chose these styles because they expressed a range of fidelity and realism in their visual hallmarks. 

The next area was the Style Explorer, shown in the middle of Figure \ref{fig:systemdesign}. Users could enter phrases about a style they wanted, after which a semantic search would retrieve relevant styles. Rationales were displayed alongside the retrieved styles to explain why the style was returned. (These rationales were the sentences scraped from GPT-3.) 

The Style Explorer also came with a set of five default styles: “abstract art”, “vector art”, “documentary photography”, “collage”, and “sketch”. These were styles that performed well with news illustrators during the formative study. The Style Explorer also automatically would try to match subjects to styles. When subjects were selected in the Style Explorer Area, these subjects were queried against the style dataset for relevant styles. At the end of the structured exploration stages, users were also given a PROMPT area to put in custom prompts of there own. In a baseline version of Opal, this area and the article area were the only features given to users.


The system was implemented in Python and Flask as well as HTML/CSS Javascript and Ajax. The text-to-image framework used was originally written by \cite{nerdyroden} and Katherine Crowson and refactored by the authors. \cite{nerdyroden, crowsondiffusion}




\section{Study 1: Exploring GPT-3 as a Source of Suggestions}
Opal automatically generates keywords, tones, and icons by prompting GPT-3. In this study, we validate our use of GPT-3 by benchmarking its results against a human standard for the tasks we included in Opal. Specifically, we address the research questions below:
\begin{itemize}
\item\textbf{RQ1.} Does GPT-3 perform as well as humans in extracting keywords and tones from article text and finding visual icons related to keywords and tones?

\item\textbf{RQ2.} Can GPT-3 reduce the mental demand of completing these tasks?

\item\textbf{RQ3.} Which of the two ways of prompting GPT-3 for artistic style recommendations performs better: direct search of styles or indirect search over style datasets?
\end{itemize}

\subsection{Methodology}
We conducted annotation studies to answer the research questions above. To prepare the data used in the study,  we randomly selected five news articles  -- three from the New York Times’ “The Year in Illustration 2021” and two from local newspaper compilations. We extracted the title, first paragraph, and last paragraph of each article. The concatenation of these elements constituted a rough summary of the article. 
 
 \subsubsection{Keywords: Task Methodology}
\begin{figure}
    \centering
    \includegraphics[width=0.5\textwidth]{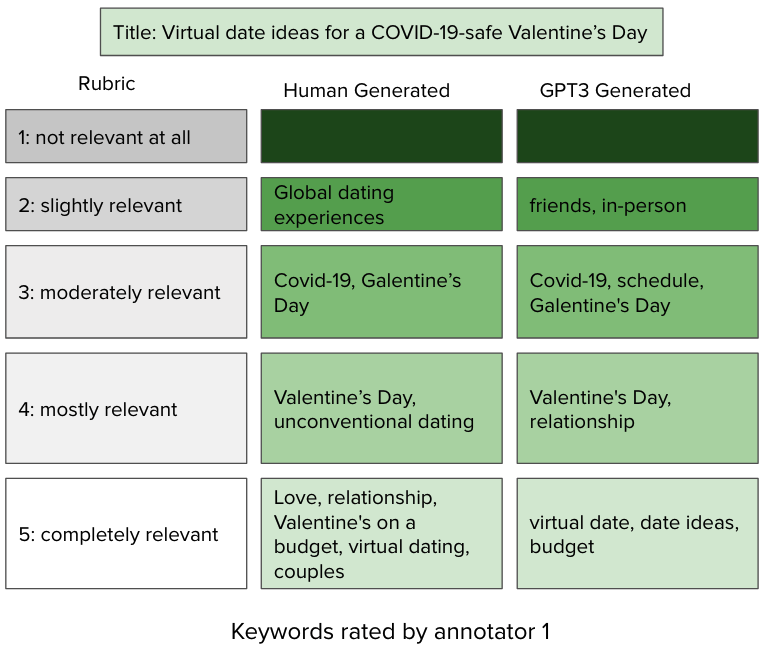}
    \caption{Examples of human-generated keywords and GPT3-generated keywords, as well how these examples were rated by annotators. The rubric quantified how relevant generated words were to the article text.}
    \Description{The image shows generated keywords grouped by ratings by annotator 1 for the article "Virtual date ideas for a COVID-19 safe Valentine's Day." The rubrics of the ratings are shown in the first column. More specifically, 1 means "not relevant at all"; 2 means "slightly relevant"; 3 means "moderately relevant"; 4 means "mostly relevant" and 5 means "completely relevant". No keywords are rated 1 in this set. Keywords being rated 2 includes human participant generated: "global dating experiences" and GPT-3 generated: "friends" and "in-person." Keywords being rated 3 includes human participant generated: "Covid-19" and "Galentine's Day" and GPT-3 generated: "Covid-19","schedule" and "Galentine's Day." Keywords being rated 4 includes human participants generated: "Valentine's Day" and "unconventional dating" and GPT-3 generated: "Valentine's Day" and "relationship". Lastly, keywords rated 5, includes human generated: "love", "relationship", "Valentine's day on a budget", "virtual dating" and "couples" and GPT-3 generated "virtual data","date ideas" and "budget".}
    \label{fig:keywordrub}
\end{figure}

We prompted GPT-3 with {\it “Give me 10 keywords associated with: [rough summary]”} to get 10 keywords for the article. A human was given the same rough summary and asked to also come up with 10 keywords. Two annotators rated GPT-3-generated and human-generated keywords in randomized order: “On a scale of 1-5, how relevant is the keyword to the article?” From five articles, 100 keywords were annotated. The rubric and example keywords are shown in Figure \ref{fig:keywordrub}.

\subsubsection{Tones: Task Methodology}

\begin{figure}
    \centering
    \includegraphics[width=0.5\textwidth]{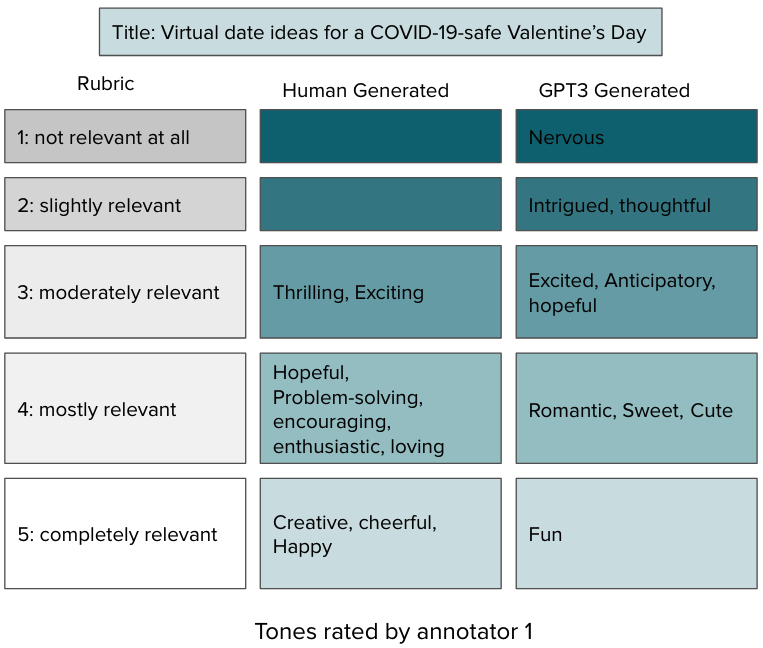}
    \caption{Examples of human-generated tones and GPT3-generated tones, as well how these examples were rated by annotators. The rubric quantified how relevant generated tones were to the article text.}
    \Description{The image shows generated tone words grouped by ratings by annotator 1 for the article "Virtual date ideas for a COVID-19 safe Valentine's Day." The rubrics of the ratings are shown in the first column. More specifically, 1 means "not relevant at all"; 2 means "slightly relevant"; 3 means "moderately relevant"; 4 means "mostly relevant" and 5 means "completely relevant". The tone that was rated 1 in this set was "nervous" generated by GPT-3. Tones being rated 2 includes GPT-3 generated: "intrigued" and "thoughtful". Keywords being rated 3 includes human participant generated: "thrilling" and "exciting" and GPT-3 generated: "excited","anticipatory" and "hopeful". Keywords being rated 4 includes human participants generated: "hopeful", "problem-solving", "encouraging", "enthusiastic" and "loving" and GPT-3 generated: "romantic", "sweet" and "cute". Lastly, keywords rated 5, includes human generated: "creative", "cheerful", and "happy" and GPT-3 generated "fun".}
    \label{fig:tonerub}
\end{figure}

Similar to the Keywords methodology, we prompted GPT-3 with {\it “Give me 10 emotions associated with: [rough summary]”} to get 10 tones. A human was given the same rough summary and asked to come up with 10 tones. Two annotators rated GPT-generated and human-generated tones in randomized order: “On a scale of 1-5, how relevant is the tone to the article?” From five articles, 100 tones were generated. The rubric and example tones are shown in Figure \ref{fig:tonerub}. 

\subsubsection{Generating Icons from Keywords}
\begin{figure}
    \centering
    \includegraphics[width=0.5\textwidth]{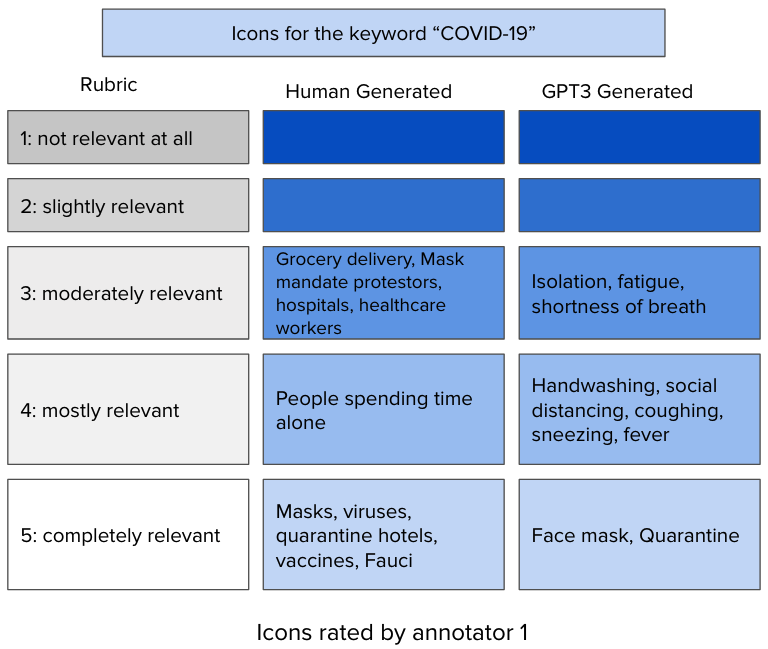}
    \caption{Examples of human-generated icons and GPT3-generated icons from selected keyword 'COVID-19', as well how these icons were rated by annotators. The rubric quantified how relevant generated icons were to the article text.}
    \Description{The image shows generated icon words grouped by ratings by annotator 1 for the the keyword "Covid-19". The rubrics of the ratings are shown in the first column. More specifically, 1 means "not relevant at all"; 2 means "slightly relevant"; 3 means "moderately relevant"; 4 means "mostly relevant" and 5 means "completely relevant". None of the icons were rated 1 or 2. Icons being rated 3 includes human participant generated: "Grocery delivery," "mask mandate protesters", "hospitals" and "healthcare workers" and GPT-3 generated: "Isolation","fatigue" and "shortness of breadth". Keywords being rated 4 includes human participants generated: "people spending time alone" and GPT-3 generated: "hand washing", "social distancing", "coughing", "sneezing" and "fever". Lastly, keywords rated 5, includes human generated: "masks", "viruses", "quarantine hotels", "vaccines" and "Fauci" and GPT-3 generated "face mask" and "quarantine".}
    \label{fig:kirub}
\end{figure}

Next, we used keyword data from the Keyword task to seed the generation of icon words. For each article, one annotator selected three of the most relevant GPT-3-generated keywords. We then prompted GPT-3: {\it “Give me 10 icons associated with [keyword]”}. We asked a human to also come up with 10 icons for those three keywords. (We chose to expand just three of the ten keywords into icons since the focus of this task was on the icon words).  Two annotators rated GPT-3-generated and human-generated icons in randomized order: “On a scale of 1-5, how relevant is the icon to the keyword?” The rubric and example icons are shown in Figure \ref{fig:kirub}. From five articles, thirteen keywords were expanded into 260 icons. (Two keywords were eliminated due to duplication.)

\subsubsection{Generating Icons from Tone}
\begin{figure}
    \centering
    \includegraphics[width=0.5\textwidth]{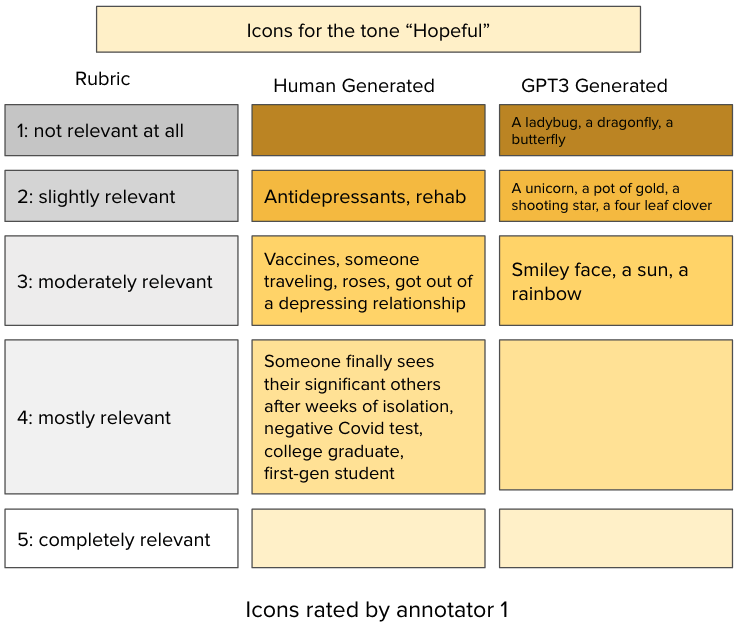}
    \caption{Examples of human-generated icons and GPT3-generated icons from selected tone 'hopeful', as well how these icons were rated by annotators. The rubric quantified how relevant generated icons were to the article text.}
    \Description{The image shows generated icons grouped by ratings by annotator 1 for the tone "hopeful". The rubrics of the ratings are shown in the first column. More specifically, 1 means "not relevant at all"; 2 means "slightly relevant"; 3 means "moderately relevant"; 4 means "mostly relevant" and 5 means "completely relevant". The icons that were rated 1 in this set include "a ladybug", "a dragonfly" and "a butterfly" generated by GPT-3. Icons being rated 2 includes human generated: "antidepressants" and "rehab" and GPT-3 generated: "a unicorn", "a pot of gold", "a shooting star" and "a four leaf clover". Keywords being rated 3 includes human participant generated: "vaccines", "someone traveling", "roses" and "got out of a depressing relationship" and GPT-3 generated: "smiley face","a sun" and "a rainbow". Keywords being rated 4 includes human participants generated: "someone finally sees their significant other after weeks of isolation", "negative Covid test", "college graduate" and "first-gen student." Lastly, no icons were being rated a 5.}
    \label{fig:tirub}
\end{figure}

We used tone data from the Tone task to seed the generation of icon words. For each article, one annotator selected three of the most relevant GPT-3-generated tones. We prompted GPT-3: {\it“Give me 10 icons associated with [tone]”}. We asked a human to also come up with 10 icons for each of those tones. Two annotators rated the GPT-3-generated and human-generated icons in randomized order: “On a scale of 1-5, how relevant is the icon to the tone?” The rubric and example data are shown in Figure \ref{fig:tirub}. From five articles, fourteen tones were expanded into 280 icons. (One tone was eliminated due to duplication.)

\subsubsection{Styles: Methodology}
\begin{figure}
    \centering
    \includegraphics[width=0.5\textwidth]{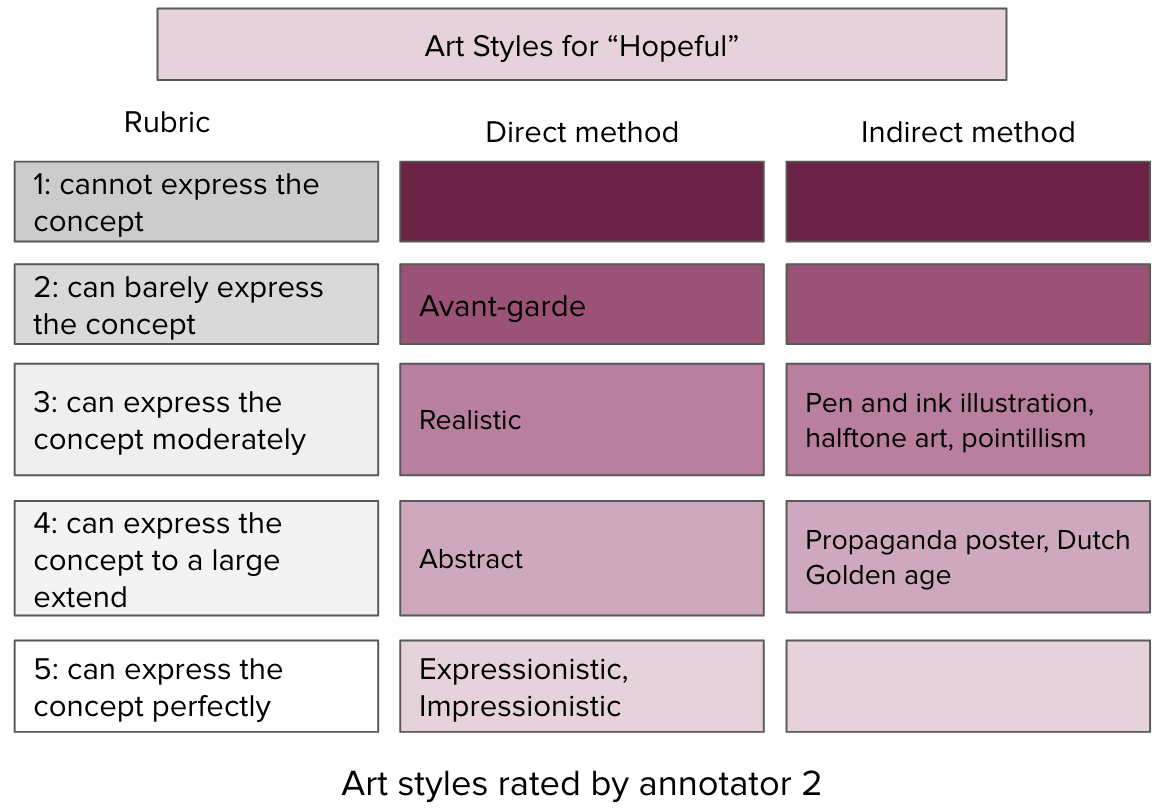}
    \caption{Examples of art styles retrieved for an article with a "hopeful" tone using two methods: 1) direct search of styles in GPT-3 or 2) indirect search of styles using GPT-3 stylistic knowledge, as well as how these styles rated on the annotator rubric.}
    \Description{The image shows generated art styles grouped by ratings by annotator 1 for the word "hopeful". The rubrics of the ratings are shown in the first column. More specifically, 1 means "the art style cannot express the concept"; 2 means "the art style can barely express the concept"; 3 means "the art style can express the concept moderately"; 4 means "the art style can express the concept to a large extend" and 5 means "the art style can express the concept perfectly". No art styles were rated 1 in this set. Art styles being rated 2 includes direct search generated: "Avant-garde". Art styles being rated 3 includes direct search generated: "Realistic" and indirect search generated: "Pen and ink illustration","Halftone art" and "Pointillism". art styles being rated 4 includes direct search generated: "Abstract" and indirect search generated: "propaganda poster" and "Dutch golden age". Lastly, art styles rated 5, includes direct search generated: "Expressionistic" and "Impressionistic".}
    \label{fig:srub}
\end{figure}

We prompted GPT-3 to generate styles in two ways. In the first way, \textit{direct search}, we directly prompted GPT-3 with: {\it“Give me 5 visual artistic styles associated with [selected keyword / tone]”}. In the second way, \textit{indirect search}, we used the 125 art styles found during co-design and prompted GPT-3 with:  {\it“Give me visual hallmarks of [style]:”}. The generated sentences then became a corpus of GPT-3 responses that we could  \textit{indirectly search} by using  semantic search with the selected keyword or tone. Five keywords and five tones from the previous GPT-3 generated keywords and tones were selected. We chose to compare these two methods of style search, because we found that GPT-3 sometimes defaulted to answering with the same set of styles when we employed direct search. Two annotators with art backgrounds rated the styles: “On a scale of 1-5, how well can the artistic style express the [keyword / tone]?” The rubric and example style data are shown in Figure \ref{fig:srub}. After each task in the aforementioned methodologies, we asked participants to answer a NASA Task Load Index questionnaire. 

\subsubsection{Recruitment of Participants}
Participants for the keyword, tone, and icon methodologies were college graduates recruited through word-of-mouth. Each of those task methodologies involved three participants: one participant to generate the human benchmark, and two participants to annotate the human-generated and GPT-3-generated words. For the last experiment on styles, which required more art background, we recruited college students in art and art history programs through word-of-mouth. All participants were compensated for \$20 per hour for however long the task and survey took.

\subsection{Results}
\subsubsection{Keywords}
For keyword ratings, we found moderate agreement between annotators (weighted Cohen's kappa=0.48). The mean rating of human-generated keywords was 4.15, and the mean rating of GPT-3 generated keywords was 3.58, as reported in Table \ref{mean1}. With a Mann-Whitney U test, we found that the this difference was significant (p-value = 0.008). For human generated keywords, 77\% were rated highly (4 or higher). For GPT-3 generated keywords, 56\% were rated highly. Although people are clearly better at extracting keywords for an article, GPT-3 still produces many relevant results.

\subsubsection{Tones}
 For tone ratings of two annotators, we report fair agreement (weighted Cohen’s Kappa= 0.28). With a Mann-Whitney U test, we found that the mean rating of human-generated tones (3.18) was significantly higher than that of GPT-3-generated tones (2.43) (to a p-value = 0.005). For human generated tones, 49\% were rated highly. For GPT-3 generated tones, 23\% were rated highly. People are better at generating tones for article, but GPT-3 still produces decent results.

\subsubsection{Generating Icons from Keywords}
For the ratings of icons from keywords, we report slight agreement (weighted Cohen’s Kappa=0.13). With a Mann-Whitney U test, we found that the mean rating of human-generated icons (4.03) was significantly higher than that of GPT-3-generated icons (3.52) (to a p-value < 0.001). For human generated icons, 71\% were rated highly. For GPT-3-generated icons, 53\% were rated highly. Again, although people are better at generating icons from keywords, but GPT-3 still produces a majority of highly relevant results.

\subsubsection{Generating Icons from Tones}
\begin{figure}
    \centering
    \includegraphics[width=0.5\textwidth]{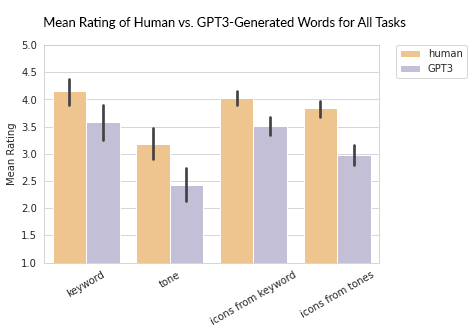}
    \caption{Mean ratings of keyword, tones and icons generated by GPT-3 and human.}
    \Description{In this bar chart figure we see bars depicting the mean rating of keywords, tones, and icons generated by GPT-3 and humans. In each pair, the human outperforms the automatic GPT-3 in terms of mean rating by approximately 0.5-1 units in rating.}
    \label{ratings1}
\end{figure}

For the ratings of icons from tones, We report moderate agreement (weighted Cohen’s Kappa=0.33). With a Mann-Whitney U test, we found that the mean rating of human generated icons (3.84) was significantly higher than that of GPT-3-generated icons (2.98) (to a p-value < 0.001).
For human generated icons, 65\% were rated highly. For GPT-3-generated icons, 37\% were rated highly. Again, people are better at coming up with relevant icons from tones, but GPT-3 still produces a good fraction of highly rated results. The results from all the tasks are summarized in Table \ref{mean1} and Figure \ref{ratings1}.

\begin{table}[]
\caption{Mean Ratings by Tasks}
\label{mean1}
\begin{tabular}{lll}
\hline
\multicolumn{1}{|l|}{}                              & \multicolumn{1}{l|}{Human} & \multicolumn{1}{l|}{GPT-3} \\ \hline
\multicolumn{1}{|l|}{Keyword}                       & \multicolumn{1}{l|}{4.15**}            & \multicolumn{1}{l|}{3.58**}           \\ \hline
\multicolumn{1}{|l|}{Tones}                         & \multicolumn{1}{l|}{3.18**}            & \multicolumn{1}{l|}{2.43**}           \\ \hline
\multicolumn{1}{|l|}{Icons generated from keywords} & \multicolumn{1}{l|}{4.03***}           & \multicolumn{1}{l|}{3.52***}          \\ \hline
\multicolumn{1}{|l|}{Icons generated from tones}    & \multicolumn{1}{l|}{3.84***}           & \multicolumn{1}{l|}{2.98***}          \\ \hline
\multicolumn{3}{l}{* P\textless{}=0.05, ** P\textless{}=0.01, ***P\textless{}=0.001}                                                
\end{tabular}
\\
\textbf{Mean ratings of how relevant sets of keywords, tones, icons generated from keywords, and icons generated from tones were to the original article.}
\end{table}

\subsubsection{NASA Task Load Index}
\begin{figure}
    \centering
    \includegraphics[width=0.5\textwidth]{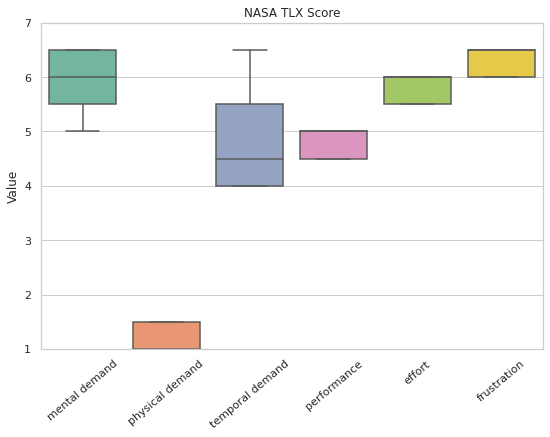}
    \caption{ Boxplots describing NASA-TLX scores which range from 1 to 7. Higher scores mean that there was worse performance or that more cognitive demand was incurred.}
    \Description{The figure shows six boxplots describing NASA-TLX scores which range from 1 to 7 for mental demand, physical demand, temporal demand, performance, effort, and frustration. Higher scores mean that there was worse performance or that more cognitive demand was incurred. Every dimension was high (meaning that the interquartile range and median were high) for this task except for physical demand.}
    \label{nasa}
\end{figure}

On average, it took the participant 25.8 minutes to complete a set of tasks for one article, and the average NASA-TLX scores are shown in Figure \ref{nasa}. The TLX scores and the time spent on these tasks indicated high levels of mental demand, effort and frustration.

\subsubsection{Generating Styles from Keywords}
For ratings of styles relevant to keywords, we report fair agreement (weighted Cohen’s Kappa = 0.28). With a Mann-Whitney U test, we found that the mean rating of direct search for styles was significantly higher than the mean rating of indirect search over the style dataset (p-value = 0.009). For direct search, 39\% were rated highly. For indirect search, 24\% were rated highly. 

\subsubsection{Generating Styles from Tones}
\begin{figure}
    \centering
    \includegraphics[width=0.5\textwidth]{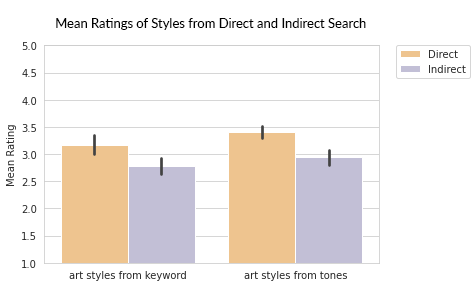}
    \caption{Mean ratings of art styles recommended by GPT-3 using the direct search method and the indirect search method.}
    \Description{In this bar chart figure we see bars depicting the mean rating of art styles generated by GPT-3 using direct and indirect search methods. In each pair, the direct search method outperforms the indirect search method in terms of mean rating by approximately 0.5-1 units in rating.}
    \label{fb}
\end{figure}

For ratings of styles relevant to tones, we report slight (weighted Cohen’s Kappa=0.1) agreement. With a Mann-Whitney U test, we found that the mean rating of styles from direct search was significantly higher than the mean rating of styles from indirect search (p-value < 0.001). For direct search , 48\% were rated highly. For indirect search, 27\% were rated highly. The results are summarized in Table \ref{mean2} and Figure \ref{fb}.

\begin{table}[]
\caption{Mean Ratings by Search Method}
\label{mean2}
\begin{tabular}{lll}
\hline
\multicolumn{1}{|l|}{}                         & \multicolumn{1}{l|}{Direct search} & \multicolumn{1}{l|}{Indirect search} \\ \hline
\multicolumn{1}{|l|}{Art styles from keywords} & \multicolumn{1}{l|}{3.18**}  & \multicolumn{1}{l|}{2.78**}   \\ \hline
\multicolumn{1}{|l|}{Art styles from tones}    & \multicolumn{1}{l|}{3.41***} & \multicolumn{1}{l|}{2.95***}  \\ \hline
\multicolumn{3}{l}{* P\textless{}=0.05, ** P\textless{}=0.01, ***P\textless{}=0.001}                         
\end{tabular}
\\
\textbf{Mean ratings of how relevant sets of styles generated from direct and indirect search were to the original article.}
\end{table}

\subsection{Conclusion}
From the annotation results, we see that GPT-3 is definitely not as good as humans at performing the keyword, tone, and icon generation tasks. However, we argue that GPT-3 can be very effective at reducing time and mental stress from our NASA-TLX results. GPT-3 can complete the same tasks a human can do in 25 minutes in under a few seconds, eliminating unnecessary mental demand, effort, and frustration. Furthermore, we found that the percentage of highly rated GPT-3 generated results was high enough to provide usable suggestions for people.\textbf{Therefore, GPT-3's competent performance on these tasks and the high mental effort that would be required by humans to generate keywords, tones, and icons validates how useful GPT-3 can be in a text-to-image system.}

For the style recommendations, we found that both methods of search had room for improvement. Although direct search was rated higher overall, direct search failed more often than indirect search at recommending styles. For example, direct search often returned random results that were not necessarily visual art styles. For example, some of the GPT-3 responses included: "social distancing art", "coronavirus art", "affordable art", "prog rock", "baroque pop" etc. Some of these were made-up art styles and others were not visual art styles but music genres. Direct search also failed to come up with consistently different recommendations. For instance, the five most frequent styles recommended by direct search were: cubism, surrealism, expressionism, abstract expressionism and neo-expressionism, which together accounted for 33\% of all recommendations. In contrast, the five most frequent results for indirect search only accounted for 17\% of all recommendations. This led us to implement indirect search in Opal. \textbf{With indirect search, by collecting datasets of stylistic knowledge from GPT-3 to search over, we could arrive at sets of style recommendations that have more variety and no chance of being made up.}

\section{Study 2: User Evaluation}

Study 1 demonstrated the validity of using GPT-3 to suggest keywords, tones, icons, and styles. Next, we wanted to understand if these techniques could structure a text-to-image system and help people achieve more usable generations for news illustration. We integrated this approach in Opal and formally investigated the following research questions for the system:
\begin{itemize}
\item\textbf{RQ1} Does Opal help users arrive at more usable generations compared to our baseline?
    
\item\textbf{RQ2} Does Opal help users arrive at a set of generations with lower cognitive load than our baseline?

\item\textbf{RQ3} Does Opal help users arrive at a set of generations with greater creative expression than our baseline?

\item\textbf{RQ4} To what extent can Opal help users create generations of usable quality as is, as visual assets, or as ideas?
\end{itemize}


\begin{figure}
    \centering
    \includegraphics[width=\columnwidth]{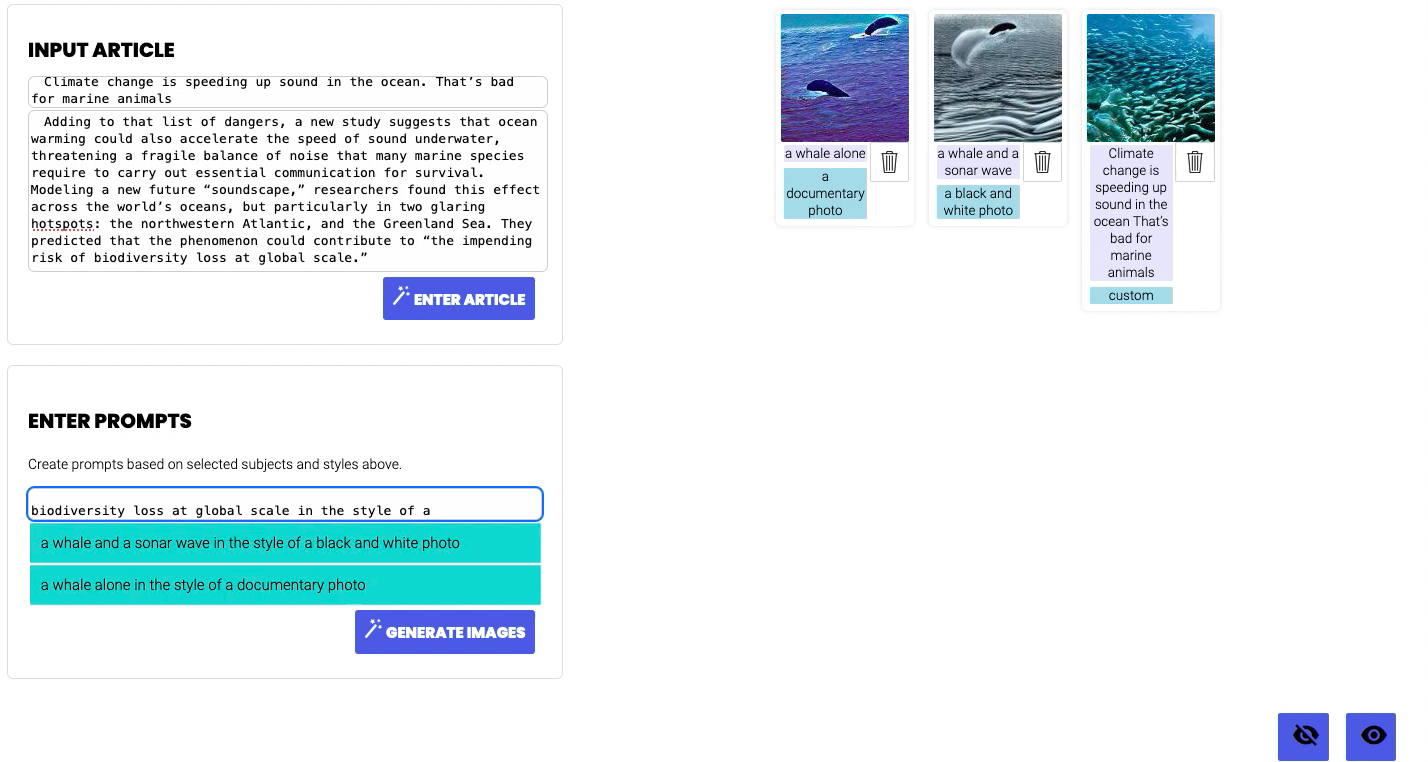}
    \caption{An ablated version of our full system for the baseline. Users were only given a textbox to prompt the text-to-image model for images.}
    \Description{This figure depicts the baseline version of our system, with a few text prompts filled in and generations populating the gallery. We used an ablated version of our full system for the baseline, leaving users only a textbox to prompt the text-to-image model for images. 
    }
    \label{fig:baseline}
\end{figure}

\subsection{Methodology}

\subsubsection{Participants.}

We recruited user study participants by emailing design organizations at our institution and going through word-of-mouth within our co-designers’ networks of design professionals. We found twelve participants, all of whom were screened to have preprofessional or professional backgrounds across the following disciplines: news illustration, fine art, user interface design, architecture, and interactive media arts. All participants performed art or design tasks as part of their job at least once a week. This study was approved by the relevant IRB.

\subsubsection{Task.}
Participants were given a task to illustrate two articles. One article was about the problem of overcrowding in national parks. Another was about healing in the time of COVID-19. These articles were chosen randomly from 62 illustrated articles listed in the New York Times “2021 The Year in Illustration” \cite{the_new_york_times_2022}.

\subsubsection{Procedure.}
We conducted a within-subjects experiments and created an ablated version of our system to use as a baseline. The ablated baseline presented the user with fields for article title and text and a simple text box for prompts. While minimal, this text-only interaction is primarily how many users report interacting with text-to-image systems in online communities \cite{R/bigsleep}, so it constituted our baseline. For each participant, the system automatically generated an image for the article title. Our baseline is depicted in Figure \ref{fig:baseline}. 

Our experiment took place online through screenshared video chat. Participants were introduced to the task and given a tour of the interface. They then generated illustrations for the two articles in two rounds and were told for which round to use Opal and which round to use the baseline. To mitigate for learning effects, we counterbalanced our conditions, giving half of the participants Opal for the first round and half of the participants the baseline for the second round. We additionally counterbalanced the order of the articles we presented. 

Users spent approximately 25 minutes with the system and the baseline each. They then spent 15 minutes post-study in a semi-structured interview. We had all users talk through the images they had generated. First, we had them eliminate generations deemed unusable. Then, we had participants talk through the remaining set of usable generations and categorize them based on if they would use them 1) as is, 2) as a visual asset that could be used with editing, or 3) as an idea. 

Often, participants found it difficult to bin generations into one category. For example, participants could find a image potentially usable both as is or as an idea. In situations like these, we rounded up to the higher degree of usability for analysis. Participants then filled out a questionnaire including NASA-TLX workload questions as well as Likert scale questions for creativity support measurement adapted from \cite{cococo}, \cite{shimizu}, and \cite{Matejka2018}. After the study, users were also given the option to be compensated for 10-15 minutes spent post-processing their visual assets, to complete the thoughts they shared during user studies. The study took about 1.5 hours and participants were compensated \$40 USD.

\subsection{Results}

\begin{figure*}
    \centering
    \includegraphics[width=\textwidth]{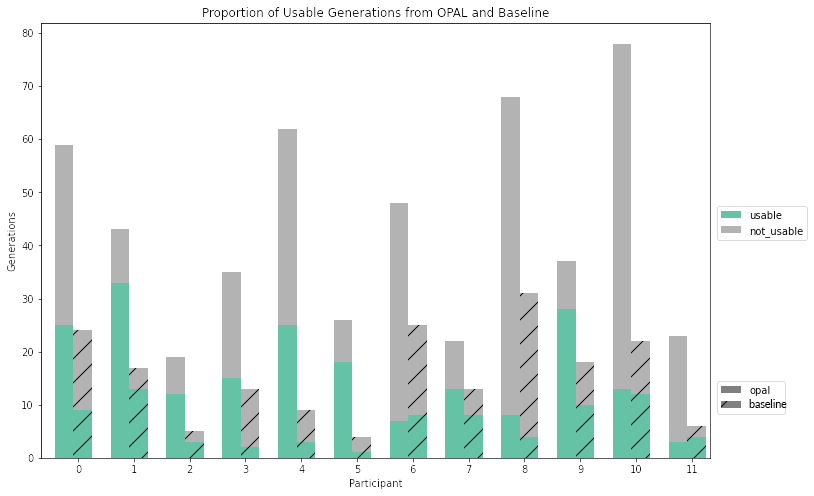}
    \caption{Across twelve participants, we found that the average number of generations created using Opal was significantly higher (2.68x) than the average number of generations created with the baseline. The number of usable generations created by Opal compared to the baseline was also significantly higher (2.28x).}
    \Description{In this figure, the proportion of usable generations are highlighted in green against gray vertical bars which highlight the number of total generations seen. Across twelve participants, we found that the average number of generations created using Opal was significantly higher (2.68x) than the average number of generations created with the baseline. The number of usable generations created by Opal compared to the baseline was also significantly higher (2.28x). }
    \label{fig:usable_proportion}
\end{figure*}

\subsubsection{RQ1: Does Opal help users arrive at a larger sets of usable generations?}

\textbf{We found that Opal significantly increased the number of generations participants generated compared to our baseline.} Across 12 participants, participants using Opal generated an average of 43 images while participants using the baseline generated an average of 16. Using a paired t-test, we found that this difference was significant. Opal increased the number of generations created by over two times ($p$=< 0.01). In Figure \ref{fig:usable_proportion} we can see this pictorially in the gray bars; the number of generations increased across all twelve participants.

We further found that Opal significantly improved the number of \textit{usable} generations compared to our baseline. With Opal, users found an average of 17 usable generations, while with the baseline, users found an average of 6 usable generations. Using a paired t-test, we again found that Opal these improvements in usable generations were significant to ($p$<0.01), meaning that Opal increased the number of usable generations by two times. In Figure \ref{fig:usable_proportion} we can see this pictorially in the green bars; the number of usable generations improved across all twelve participants. \textbf{Given these results, we conclude that Opal does allow users to arrive at a larger set of usable generations within the same amount of time.}

\subsubsection{RQ2: Does Opal help users with cognitive load?}

\begin{figure}
    \centering
    \includegraphics[width=0.5\textwidth]{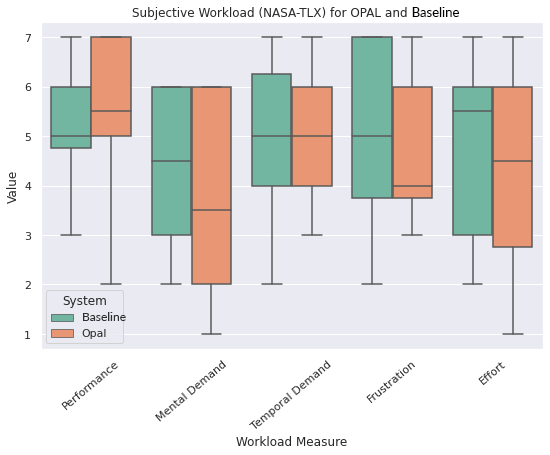}
    \caption{Boxplots describing NASA-TLX subjective workload measures which range from 1 to 7. Higher scores mean that there was better performance or that less cognitive demand was incurred. We did not find significant difference between Opal and the baseline in terms of any dimension. }
    \Description{
    Boxplots describing NASA-TLX subjective workload measures which range from 1 to 7. Higher scores mean that there was better performance or that less cognitive demand was incurred. We did not find significant difference between Opal and the baseline in terms of any dimension.
    }
    \label{fig:nasatlx}
\end{figure}

\begin{figure}
    \centering
    \includegraphics[width=0.5\textwidth]{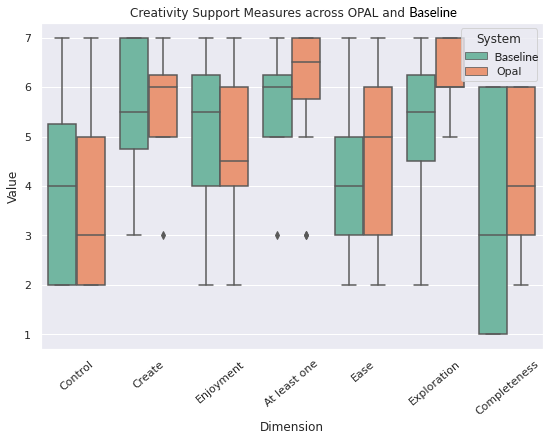}
    \caption{Boxplots describing creativity support measures which ranged from 1 to 7 (disagree to agree). Higher scores indicated agreement that the system fulfilled them in these measures: user control, user ability to create novel things, enjoyment, ability to create at least one design, ease of use, ability to explore, and exploration completeness. We found significant difference ($\alpha=0.05$) in terms of exploration. Opal helped users better explore the space of potential designs.}
    \Description{Boxplots describing creativity support measures which ranged from 1 to 7 (disagree to agree). Higher scores indicated agreement that the system fulfilled them in these measures: user control, user ability to create novel things, enjoyment, ability to create at least one design, ease of use, ability to explore, and exploration completeness. We found significant difference ($\alpha=0.05$) in terms of exploration in that Opal helped users better explore the space of potential designs. The differences in the boxplots for the other dimensions was insignificant.}
    \label{fig:creativity}
\end{figure}

We found that Opal was rated highly in terms of performance. On average, Opal was given a 5.45 out of 7 in terms of performance (a NASA-TLX measure), with half the participants giving Opal above a 6 out of 7 in terms of performance. In testing for statistical significance using paired t-tests, we found that across the NASA-TLX measures, the differences between the averages of Opal and baseline on performance, temporal demand, mental demand, effort, and frustration were insignificant. However in studying Figure \ref{fig:nasatlx}, we can see more nuances in the distributions for each measure. We see that the median performance for Opal compared to the baseline is higher. In terms of mental demand, we also see a very wide spread in the mental demand measure for Opal, captured in the boxplot as a high interquartile range and in the standard deviation = 1.9. For participants, Opal performed near neutrally in terms of frustration (mean=4.36) and effort (mean=4), which was lower than the frustration (mean=4.81) and effort (mean=4.90) scores for the baseline. \textbf{From these results, we cannot conclude that Opal has lower cognitive load than the baseline.} 

\subsubsection{RQ3: Does Opal help users arrive at a set of generations with greater creative support?}

We analyzed the creativity support measures and found through paired t-tests that Opal was significantly better at helping users in terms of exploration ($\alpha$=0.05) compared to the baseline system. Ten of twelve of the users rated \textit{Exploration} highly at around 6 or 7, which is also illustrated in the boxplot for exploration in Figure \ref{fig:creativity}. In the other measures (control, ability to create novel generations, enjoyment, ease, and completeness) we found no significant difference between Opal and the baseline. \textbf{From these results, we conclude that Opal does increase creativity support, but only in the dimension of \textit{exploration.}}

\subsubsection{RQ4: To what extent can Opal help users create generations of usable quality: as is, as visual assets, or as ideas?}

\begin{figure*}
    \centering
    \includegraphics[width=\textwidth]{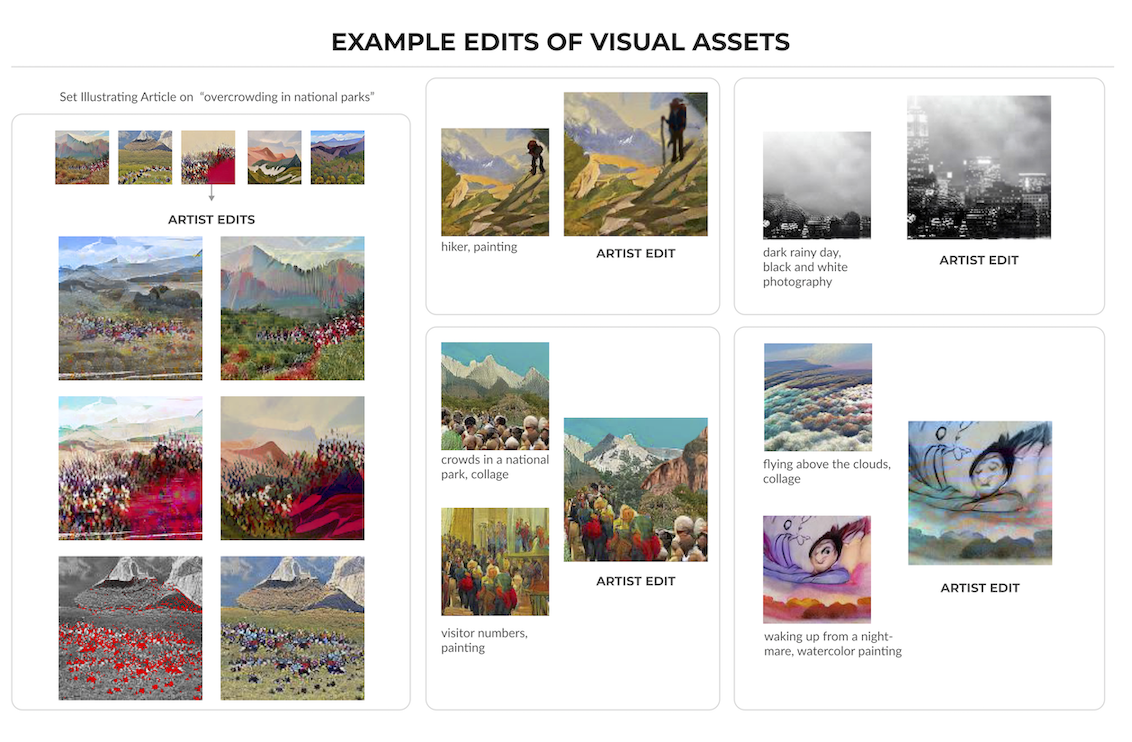}
    \caption{Multiple examples of edits different participants made quickly (<10-15 minutes) to illustrate the concepts of how they would use generations as visual assets. Each rounded rectangle shows an illustration or set of illustrations edited by a participant.}
    \Description{
    Figure has multiple regions showing how different artists made different edits from the generations as visual assets. In the leftmost column, an artist took 5 generations and remixed them with image effects like collaging and desaturation. They made 6 artist edits for the article they were given on 'overcrowding in national parks'. Another artist in the middle top column made a simple edit to correct the way the subject was represented in a generation of 'hiker in the style of painting'. In the middle bottom column, the artist collaged the two generations 'visitor numbers in the style of painting' and 'crowds in a national park in the style of collage' into one artist edit. In the right top column, the artist added some imagery in the background of a black and white photo for a richer composition. In the bottom right column, the artist collaged two generations "waking up from a nightmare in the style of watercolor painting" and "flying above the clouds in the style of collage", such that the former was used as the foreground and the latter was used as the background.
    }
    \label{fig:visual_assets}
\end{figure*}

Participants tended to only choose a few generations that they would take for use as is, if any at all. However, participants tended to regard these generations more positively as visual assets or ideas. For Opal, 10 of 12 participants were able to come up with at least one generation they could use as a visual asset. In the baseline, only 7 participants were able to come up with at least one generation for a visual asset. Illustrations made by participants who engaged with the generations as visual assets are displayed in Figure \ref{fig:visual_assets}. For example, one participant created an artist edit to show overcrowding in national parks by creating a visual blend of a generation prompt "visitor numbers in the style of painting" and "crowds in a national park in the style of collage". By masking and overlaying the generations as visual assets, the artifacts of each image were compensated for by the other. Lastly, in terms of ideas, for Opal, 11 were able to come up with at least one generation that gave them an idea. For the baseline, 10 participants came up with at least one generation they would use as an idea.

As users pared their galleries into sets of usable generations, we observed reasons why participants would find some generations not usable. Aside from the obviously blank generations the AI would occasionally return (when the AI model failed to optimize towards the text prompt), participants noted that artifacts in the generations, distortions in animal or people figures, and uncanny compositions made them find a generation unusable.

\begin{figure*}
    \centering
    \includegraphics[width=\textwidth]{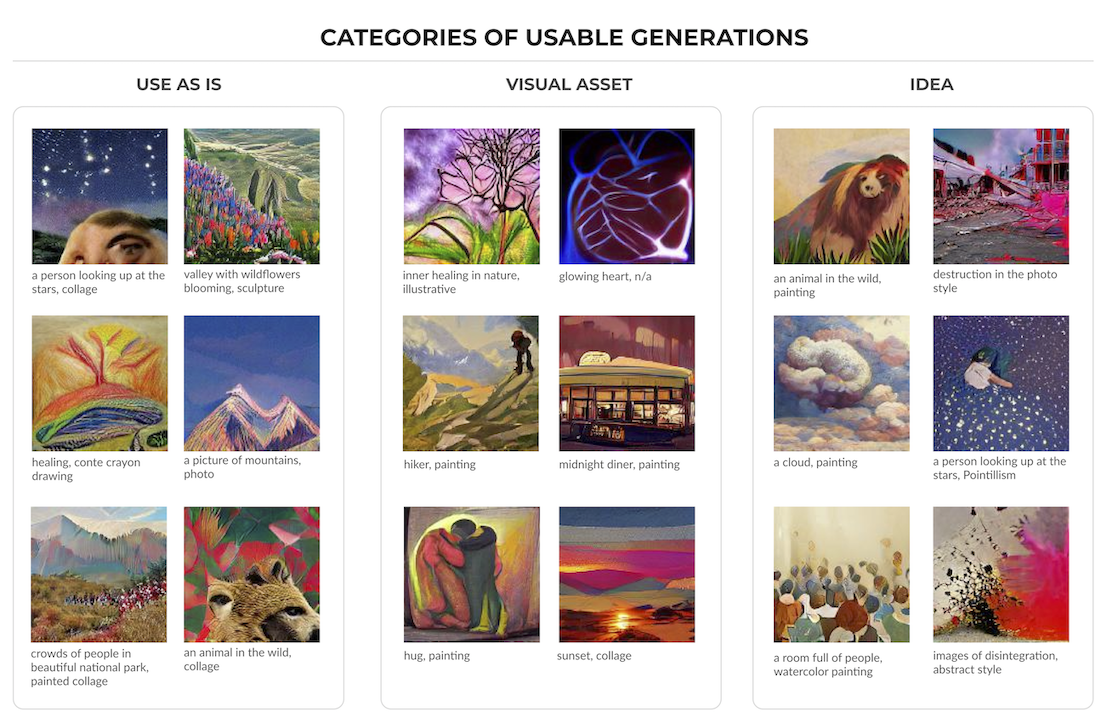}
    \caption{Participants found usable generations in all three of the categories we defined for usability: use as is, use as a visual asset, and use as a conceptual idea. Examples are shown above, with the subject and style delimited by a comma. }
    \Description{Figure has three columns showing the different categories of usable generations. Each column has six example generations and their text prompt underneath. Use as is was the left most column. This represents generations that were high quality aesthetic outcomes that could be used with little to no image processing. The middle column shows visual assets, which represented the category where artists could use the generations as base material. The last rightmost column was the idea column. These generations spawned ideas for the artists, even if they did not explicitly use the generated image as design material.}
    \label{fig:example_categorizations}
    
\end{figure*}

\subsubsection{General Qualitative Findings}

We found that participants were able to produce so many generations with Opal because when presented with suggestions, they were willing to try more. P7 stated that in Opal they wanted to give each angle of exploration “a fair shot”, whereas with the baseline, the less feedback the baseline gave them, the less willing they were to try. When P7 used the baseline, they generated four images and stopped, finding the first two generations unusable and the others only lukewarm as concepts. When P7 used Opal, they considered keywords such as "Sierra Nevada", "Conservation", "national parks", and "an animal in the wild" and then drilled down into icons that could visualize national parks, which ranged from detailed subjects ("a person with their hand in their air", "an animal in the wild" ) to named entities ("Acadia National Park"). They found inspiration in a generation of "an animal in the wild in the style of painting" and found a generation of "an animal in the wild in the style of collage" usable as is. This willingness to try extended to other features as well like the STYLE EXPLORER. Participants enjoyed trying diverse styles such as conte crayon drawing, decollage, or double exposure photography --- even if they had no exposure to them prior. 

Participants unanimously reported that keywords were helpful for anchoring their exploration. However, participants noted that sometimes tones could be less helpful, because the stage could take them in divergent directions from the ones they had just arrived at with keywords. For example, for P4, after having generated concrete images such as "trees", "a picture of a campground", and "a megaphone", they noted they were being pulled out of that line of focus to generate "serenity", "wonder", and other tone-related icons. 

Participants also reported that the amount of choices returned when tones and keywords were expanded into icons could at times be overwhelming, repetitive, or overspecific. To quote P9, \textit{“Keywords give me generic images. I like how these [icons] start to get more specific, more rich conceptually. Some of these were a little bit too predictable, like a broken heart or a wilted flower, I feel like you see these images a lot.}

Another theme of note is that Opal encouraged reaction over intention for some participants. P1 commented,  \textit{“I’m reacting more to what it’s generating than me proactively putting in things. It’s nice for generating ideas, it’s good for cherry picking. Here [Opal] spits out so many, I can just do this forever until I get that one ideal one. The other one [baseline] took a lot more effort in terms of generating an image.”} Similarly, P6 acknowledged that they were more interested in the outcome than the prompts driving the process. \textit{“I wish it just did everything for me. I really don’t care if I chose the right keyword, but I wanted the right outcome.”} In contrast, P10 highlighted that the baseline allowed them to \textit{ "have control of what I am searching. I can come up with keywords on my own, which gives me more time to think about what I want to see more intentionally."}

Some participants mentioned that the way they viewed the generations was influenced by the amount of effort they put in. P1 commented that because the baseline took deliberate search on their part for a prompt, they felt less likely to edit the images and more likely to take them as is. P12 similarly mentioned that the text-to-image AI seemed to be giving them "solutions” and that they did not feel like they would significantly modify the images because "it would interrupt the aesthetic of the generation". 

We note that when participants were presented with the baseline first, they tended to experiment with the prompts to build a concept of what the system could or could not do, even if the experimentation was tangential to their goal news graphic. Experimentation was highly individual and curiosity-based. P12 wanted to understand if the system could handle abstraction–\textit{“I tried economics, conservation… abstract ideas to see if the AI could generate anything that could be used stylistically more than just subject matter.} P6 attempted different linguistic arrangements within the prompt, at first trying to convey photographic qualities (i.e. “close up of a hiker in the style of photography), and then different prepositions (an elk centered in front of a mountain the style of”, and additionally plurality (i.e. “two hikers”). P5 replaced the word “a person” with “woman”, wondering if the model would respond better if the subject was gendered. From experimentation, participants would make assumptions about model capabilities. Some (P1, P2, P3) became more reluctant to try subjects that were nonhuman, figurative, and animal things. Others (P11, P12) presumed after just a few generations that the model might be better at scenic images or "high texture ones" that possess an "uncontrolled" quality often seen in nature. 

The three news illustrators within our participant group were also receptive to Opal. Two of the news illustrators in their thinkaloud process mentioned that even if they had a concept in mind at the start that they were not able to directly input into a text box, they were able to work with the generations and build on top of their mental image nonetheless. For example, P11, an editorial illustrator for a community newspaper said that they had a mental image from the start of a glowing heart”. They began putting in prompts related to "glowing hearts", and eventually converged on some concepts.\textit{ "I got more clarity on what the background could look like, because to do with nature. and this gave me a color palette, some texture, some more concrete ways to represent that. So then from here I could not only use these images but find ones that are similar to it and look up fields that look similar to that one to use in that image."}

News illustrators P11 and P8 both mentioned that one of the greatest advantages to text-to-image systems is less copyright concerns. P8 mentioned that as opposed to browsing the internet for inspiration and worrying about copyright infringement (i.e. accidentally taking inspiration from other people's illustrations), they could freely use the generations from Opal in any way. Both referred to Opal as a great resource for generating reference photos that could be traced over.


Overall, participants enjoyed the tailored, cumulative, and yet explorative nature of Opal. P12 in particular enjoyed Opal, saying \textit{"I think it was like a pipeline. You start with an idea and you develop it more. The other one [baseline] felt like going to ground zero each time. It didn’t feel like the ideas were building on each other that much. You were still in the same zone, you were still dealing with the same thing, it was changing but it didn’t seem like it was accumulating as much.”}

\section{Discussion}

We center our discussion around the following generalizations from our results: 1) Generating more images gives users better outcomes and helps them better understand AI capabilities 2) Multiple general-purpose AI models can integrate into one workflow 3) Generative AI should augment rather than replace designers. 

\subsection{Generating more images gives users better outcomes and helps them understand AI capabilities}

One of our main quantitative findings was that Opal users found over twice the number of usable outcomes than users without it. This is consistent with findings from ideation literature stating that “more is better” when it comes to ideas. Generating more allows users to extensively explore the space of design solutions and better understand that they have the freedom to experiment with a near infinite number of visual concepts. We encourage this in Opal by enabling users to efficiently generate by embedding “default settings” within our system such as a prompt format (“subject in the style of”) and default styles (photo, painting, collage)---helping them efficiently reach better outcomes.

Additionally, when users see more generations, they are also given more visual evidence of what a generative AI is capable of. This helps users craft a mental model of the AI. Gero et. al. \cite{gero2020} found that it is important for users during interaction with AI to develop mental concepts of what range of knowledge an AI can display. We note that when users were first introduced to the baseline during our user study, they often experimented with prompts, just to get a quick impression of what the model could and could not do, even if those prompts were tangential to their goal illustration. Because state-of-the-art models are so large, we expect that users will need to continuously refine their mental model of what these AI are capable of even after months of use. This is evident in the communities evolving on social media that explore AI and share the design patterns they have discovered. Thus, encouraging exploration not only helps users achieve better outcomes on their current work but also improves their work over time as users become more knowledgeable. 

When users pick from a large gallery of options, they also get greater agency in the form of choice. It is difficult to control a text-to-image generation outcome, because prompts will always underspecify the image and leave large parts of the composition, details, and colors up to the AI. With something as novel and stochastic as text-to-image generation, it is potentially better to allow users to choose rather than to control. It is comparable to taking a gardener’s approach as opposed to an architect’s approach. Gardeners seed but do not control the outcome; likewise in Opal, users curated from a garden of results that they could prune and branch off from. Architects blueprint the final generation and define constraints, but the fulfillment of constraints is still hard to guarantee from AI systems. Allowing the user to be a gardener rather than the architect could be the right approach for artists, who do not have to play tug-of-war with the model and always get the final say on what is visually desirable. Future systems should consider what dimensions are best to prune on. 

 In future systems where AI generations are even faster to produce, users can greatly benefit from seeing even more generations so long as they are not burdened by the creation or curation of those options. Our results measuring cognitive load illustrate an important trade-off: if users are presented with too many prompt suggestions without the ability to control for how generic or specific they are, the add added to their cognitive load and induce choice paralysis. Reducing cognitive load could be as simple as controlling the number of suggestions returned or as complex as crafting the prompt engineering to be more contextual and user dependent. The most important thing is to prevent users from seeing large amounts of random or unsatisfying results. Overall, generating more gives users better outcomes when they are exploring and understanding generative capabilities.
 
 \subsection{Multiple general-purpose AI models can integrate into one workflow}
 
 In Opal, we created a workflow that composed a language generation model (GPT-3) with a multimodal image generation model (VQGAN+CLIP). These are both large pretrained models that have different computational sensibilities. While text-to-image models can reach a near infinite number of visual concepts through text, users can find it taxing to come up with the right visual language on their own. Having a language generation model to complement the text-to-image model in Opal helped because language generation models are capable of returning any number of diverse linguistic trajectories at once. As we found in Study 1, it is remarkably challenging and taxing for a human to come up with a large, diverse set of associations. Even though GPT-3 could not beat out the human benchmark in terms of performance, what GPT-3 can do is \textit{significantly mitigate human effort} by suggesting concepts. 
 
 Having access to many different possibilities helped mitigate the odds of a user running into a poor quality text-to-image generation. The magnitude of possible text-to-image generations was contended well with the magnitude of possible language model generations---the strengths of the language model addressed a problem of the image generation model, and the image model extended the capabilities of the language generation model into the image domain.

 
In composing GPT-3 with VQGAN+CLIP, we were able to take article text, expand it into networks of concepts, and translate those concepts into prompts of rich visual language. When multiple models are integrated together into one workflow, they must be able to find common conceptual spaces such that the output of one can easily be transformed into the input of another. In Opal, this conceptual space was visual language: concrete subjects and styles that were salient with color and artistic techniques. Other examples of conceptual spaces could be the multimodal spaces between language and 3D shape or emotion and music.  

 \subsection{Generative AI should augment rather than replace designers}
 We found in our user study that news illustrators were receptive to Opal and could greatly benefit from the computational assistance of text-to-image AI. Participants were able to find strong use cases in their generations, taking them for use as is, treating them as design material / reference images, or simply inspiration. Additionally, participants touched upon some of the advantages text-to-image models had over traditional tools like Google Images or Pinterest. As opposed to browsing the internet for inspiration, where one could run into potential concerns of copyright infringement or accidentally draw too much inspiration from others illustrations, our participants felt they could freely use the generations from Opal in any way. Text-to-image systems also have the advantage that they can generate images into galleries infinitely but in a curated and mostly controlled manner. For example, with Google Images or Pinterest, users are often overwhelmingly returned with millions of tangential results, whereas text-to-image systems can return more intentional sets.

 
A valid concern with any text-to-image model is that if it is sufficiently strong at generating high-fidelity images, artists and designers will be replaced. However, we found that artistic knowledge was still necessary at every point of the process, from knowing the right artistic concepts and vocabulary to compose a prompt to having the editorial eye to select usable generations. Text-to-image AI is insufficient alone; in co-design, we tried generating images directly from news headlines, but they almost always produced results of poor quality and relevance. Having artists not only in the loop but driving the process with their insight into what constitutes strong visual language is essential.


Lastly, these models draw from old traditions of art and design, so researchers and practitioners should focus on how they can innovate text-to-image generations as a new format of art rather than a medium that regurgitates the past. For all these reasons, the focus of generative AI deployment should be to augment rather than to replace human creative expertise.

\section{Limitations and Future Work}

One limitation of our work was that we supported the news illustration process with a novel technology that departed from the usual process in a very significant way. Four participants reported that this process was an inversion of what certain parts of their illustration or design process were like. While we did structure Opal through co-design with news illustrators, involving a focus on keywords and visual concept exploration, our user interface and steering controls for the AI generation were primarily text-based. Many participants mentioned that image creation has fundamentally been about interacting with the image through direct manipulation, establishing composition spatially as opposed to through language. Therefore, a limitation within our system was that we did not give users the ability to work more off of images and other processes they may have been more traditionally used to. While we had tried to implement ways of letting users pass in image prompts, which is possible with these technologies \cite{nerdyroden}, we found that image prompts added another layer of stochasticity that would have compounded the lack of user control a user could feel. 


Additionally, we could have refined our system to be more specific to different forms of news, such as sports, lifestyle, or political news illustration. Furthermore, we acknowledge that fine-tuning on certain classes of images could have improved the coherence of images in those classes. We leave domain specificity and other features relevant to fine-tuning to future work.




\section{Conclusion}
Text-to-image models give users the ability to prompt AI for images using visual language. We explore applications of this novel technology for news illustration with Opal, a system that guides users through a structured search for visual concepts and provides pipelines allowing users to illustrate based on an article’s tone, keywords, and intended illustration style. Our evaluation shows that Opal efficiently generates diverse sets of editorial illustrations, graphic assets, and concept ideas. Users with Opal were two times more efficient at generation and generated two times more usable results than users without. We conclude on a discussion of how structured and rapid exploration can help users better understand the capabilities of human AI co-creative systems.

\begin{acks}
This work is supported by NSF DGE - 1644869.
\end{acks}

\bibliographystyle{ACM-Reference-Format}
\bibliography{citations}


\begin{thebibliography}{54}


\ifx \showCODEN    \undefined \def \showCODEN     #1{\unskip}     \fi
\ifx \showDOI      \undefined \def \showDOI       #1{#1}\fi
\ifx \showISBNx    \undefined \def \showISBNx     #1{\unskip}     \fi
\ifx \showISBNxiii \undefined \def \showISBNxiii  #1{\unskip}     \fi
\ifx \showISSN     \undefined \def \showISSN      #1{\unskip}     \fi
\ifx \showLCCN     \undefined \def \showLCCN      #1{\unskip}     \fi
\ifx \shownote     \undefined \def \shownote      #1{#1}          \fi
\ifx \showarticletitle \undefined \def \showarticletitle #1{#1}   \fi
\ifx \showURL      \undefined \def \showURL       {\relax}        \fi
\providecommand\bibfield[2]{#2}
\providecommand\bibinfo[2]{#2}
\providecommand\natexlab[1]{#1}
\providecommand\showeprint[2][]{arXiv:#2}

\bibitem[\protect\citeauthoryear{??}{met}{[n.d.]}]%
        {meta_ai}
 \bibinfo{year}{[n.d.]}\natexlab{}.
\newblock \bibinfo{title}{Greater Creative Control for AI image generation}.
\newblock
\newblock
\urldef\tempurl%
\url{https://ai.facebook.com/blog/greater-creative-control-for-ai-image-generation/}
\showURL{%
\tempurl}


\bibitem[\protect\citeauthoryear{??}{coc}{2020}]%
        {cococo}
 \bibinfo{year}{2020}\natexlab{}.
\newblock \bibinfo{booktitle}{\emph{CHI '20: Proceedings of the 2020 CHI
  Conference on Human Factors in Computing Systems}} (Honolulu, HI, USA).
  \bibinfo{publisher}{Association for Computing Machinery},
  \bibinfo{address}{New York, NY, USA}.
\newblock
\showISBNx{9781450367080}


\bibitem[\protect\citeauthoryear{??}{the}{2022}]%
        {the_new_york_times_2022}
 \bibinfo{year}{2022}\natexlab{}.
\newblock \bibinfo{title}{The Year in Illustration}.
\newblock
\newblock
\urldef\tempurl%
\url{https://www.nytimes.com/interactive/2022/01/05/multimedia/year-best-illustration-2021.html}
\showURL{%
\tempurl}


\bibitem[\protect\citeauthoryear{Adverb}{Adverb}{2021}]%
        {adverb}
\bibfield{author}{\bibinfo{person}{Adverb}.} \bibinfo{year}{2021}\natexlab{}.
\newblock \bibinfo{title}{Advadnoun}.
\newblock
\newblock
\urldef\tempurl%
\url{https://twitter.com/advadnoun}
\showURL{%
\tempurl}


\bibitem[\protect\citeauthoryear{Aggarwal and Parikh}{Aggarwal and
  Parikh}{2020}]%
        {aggarwal2020neurosymbolic}
\bibfield{author}{\bibinfo{person}{Gunjan Aggarwal} {and} \bibinfo{person}{Devi
  Parikh}.} \bibinfo{year}{2020}\natexlab{}.
\newblock \bibinfo{title}{Neuro-Symbolic Generative Art: A Preliminary Study}.
\newblock
\newblock
\showeprint[arxiv]{2007.02171}~[cs.AI]


\bibitem[\protect\citeauthoryear{Augustin, Leder, Hutzler, and Carbon}{Augustin
  et~al\mbox{.}}{2008}]%
        {augustin}
\bibfield{author}{\bibinfo{person}{M Augustin}, \bibinfo{person}{Helmut Leder},
  \bibinfo{person}{Florian Hutzler}, {and} \bibinfo{person}{Claus-Christian
  Carbon}.} \bibinfo{year}{2008}\natexlab{}.
\newblock \showarticletitle{Style follows content: On the microgenesis of art
  perception}.
\newblock \bibinfo{journal}{\emph{Acta psychologica}}  \bibinfo{volume}{128}
  (\bibinfo{date}{06} \bibinfo{year}{2008}), \bibinfo{pages}{127--38}.
\newblock
\urldef\tempurl%
\url{https://doi.org/10.1016/j.actpsy.2007.11.006}
\showDOI{\tempurl}


\bibitem[\protect\citeauthoryear{Branwen}{Branwen}{2020}]%
        {gwern_2020}
\bibfield{author}{\bibinfo{person}{Gwern Branwen}.}
  \bibinfo{year}{2020}\natexlab{}.
\newblock \bibinfo{title}{Gpt-3 creative fiction}.
\newblock
\newblock
\urldef\tempurl%
\url{https://www.gwern.net/GPT-3}
\showURL{%
\tempurl}


\bibitem[\protect\citeauthoryear{Brown, Mann, Ryder, Subbiah, Kaplan, Dhariwal,
  Neelakantan, Shyam, Sastry, Askell, Agarwal, Herbert-Voss, Krueger, Henighan,
  Child, Ramesh, Ziegler, Wu, Winter, Hesse, Chen, Sigler, Litwin, Gray, Chess,
  Clark, Berner, McCandlish, Radford, Sutskever, and Amodei}{Brown
  et~al\mbox{.}}{2020}]%
        {gpt3}
\bibfield{author}{\bibinfo{person}{Tom~B. Brown}, \bibinfo{person}{Benjamin
  Mann}, \bibinfo{person}{Nick Ryder}, \bibinfo{person}{Melanie Subbiah},
  \bibinfo{person}{Jared Kaplan}, \bibinfo{person}{Prafulla Dhariwal},
  \bibinfo{person}{Arvind Neelakantan}, \bibinfo{person}{Pranav Shyam},
  \bibinfo{person}{Girish Sastry}, \bibinfo{person}{Amanda Askell},
  \bibinfo{person}{Sandhini Agarwal}, \bibinfo{person}{Ariel Herbert-Voss},
  \bibinfo{person}{Gretchen Krueger}, \bibinfo{person}{Tom Henighan},
  \bibinfo{person}{Rewon Child}, \bibinfo{person}{Aditya Ramesh},
  \bibinfo{person}{Daniel~M. Ziegler}, \bibinfo{person}{Jeffrey Wu},
  \bibinfo{person}{Clemens Winter}, \bibinfo{person}{Christopher Hesse},
  \bibinfo{person}{Mark Chen}, \bibinfo{person}{Eric Sigler},
  \bibinfo{person}{Mateusz Litwin}, \bibinfo{person}{Scott Gray},
  \bibinfo{person}{Benjamin Chess}, \bibinfo{person}{Jack Clark},
  \bibinfo{person}{Christopher Berner}, \bibinfo{person}{Sam McCandlish},
  \bibinfo{person}{Alec Radford}, \bibinfo{person}{Ilya Sutskever}, {and}
  \bibinfo{person}{Dario Amodei}.} \bibinfo{year}{2020}\natexlab{}.
\newblock \bibinfo{title}{Language Models are Few-Shot Learners}.
\newblock
\newblock
\urldef\tempurl%
\url{https://doi.org/10.48550/ARXIV.2005.14165}
\showDOI{\tempurl}


\bibitem[\protect\citeauthoryear{Calderwood, Qiu, Gero, and Chilton}{Calderwood
  et~al\mbox{.}}{2018}]%
        {calderwood_how_2018}
\bibfield{author}{\bibinfo{person}{Alex Calderwood}, \bibinfo{person}{Vivian
  Qiu}, \bibinfo{person}{Katy~Ilonka Gero}, {and} \bibinfo{person}{Lydia~B
  Chilton}.} \bibinfo{year}{2018}\natexlab{}.
\newblock \showarticletitle{How {Novelists} {Use} {Generative} {Language}
  {Models}: {An} {Exploratory} {User} {Study}}. In
  \bibinfo{booktitle}{\emph{23rd {International} {Conference} on {Intelligent}
  {User} {Interfaces}}}. \bibinfo{publisher}{ACM}.
\newblock
\showISBNx{978-1-4503-4945-1}


\bibitem[\protect\citeauthoryear{Chakrabarty, Zhang, Muresan, and
  Peng}{Chakrabarty et~al\mbox{.}}{2021}]%
        {chakrabarty_mermaid_2021}
\bibfield{author}{\bibinfo{person}{Tuhin Chakrabarty}, \bibinfo{person}{Xurui
  Zhang}, \bibinfo{person}{Smaranda Muresan}, {and} \bibinfo{person}{Nanyun
  Peng}.} \bibinfo{year}{2021}\natexlab{}.
\newblock \showarticletitle{{MERMAID}: {Metaphor} {Generation} with {Symbolism}
  and {Discriminative} {Decoding}}. In \bibinfo{booktitle}{\emph{Proceedings of
  the 2021 {Conference} of the {North} {American} {Chapter} of the
  {Association} for {Computational} {Linguistics}: {Human} {Language}
  {Technologies}}}. \bibinfo{publisher}{Association for Computational
  Linguistics}, \bibinfo{address}{Online}, \bibinfo{pages}{4250--4261}.
\newblock
\urldef\tempurl%
\url{https://doi.org/10.18653/v1/2021.naacl-main.336}
\showDOI{\tempurl}


\bibitem[\protect\citeauthoryear{Chang, Eric, Savva, and Manning}{Chang
  et~al\mbox{.}}{[n.d.]}]%
        {sceneseer}
\bibfield{author}{\bibinfo{person}{Angel~X. Chang}, \bibinfo{person}{Mihail
  Eric}, \bibinfo{person}{Manolis Savva}, {and} \bibinfo{person}{Christopher~D.
  Manning}.} \bibinfo{year}{[n.d.]}\natexlab{}.
\newblock \bibinfo{title}{SceneSeer: 3D Scene Design with Natural Language}.
\newblock
\newblock
\urldef\tempurl%
\url{https://doi.org/10.48550/ARXIV.1703.00050}
\showDOI{\tempurl}


\bibitem[\protect\citeauthoryear{Chaudhuri, Kalogerakis, Giguere, and
  Funkhouser}{Chaudhuri et~al\mbox{.}}{2013}]%
        {Chaudhuri:2013:ACC}
\bibfield{author}{\bibinfo{person}{Siddhartha Chaudhuri},
  \bibinfo{person}{Evangelos Kalogerakis}, \bibinfo{person}{Stephen Giguere},
  {and} \bibinfo{person}{Thomas Funkhouser}.} \bibinfo{year}{2013}\natexlab{}.
\newblock \showarticletitle{{AttribIt}: Content Creation with Semantic
  Attributes}.
\newblock \bibinfo{journal}{\emph{ACM Symposium on User Interface Software and
  Technology (UIST)}} (\bibinfo{date}{Oct.} \bibinfo{year}{2013}).
\newblock


\bibitem[\protect\citeauthoryear{Cho, Zala, and Bansal}{Cho
  et~al\mbox{.}}{2022}]%
        {dall-eval}
\bibfield{author}{\bibinfo{person}{Jaemin Cho}, \bibinfo{person}{Abhay Zala},
  {and} \bibinfo{person}{Mohit Bansal}.} \bibinfo{year}{2022}\natexlab{}.
\newblock \bibinfo{title}{DALL-Eval: Probing the Reasoning Skills and Social
  Biases of Text-to-Image Generative Transformers}.
\newblock
\newblock
\urldef\tempurl%
\url{https://doi.org/10.48550/ARXIV.2202.04053}
\showDOI{\tempurl}


\bibitem[\protect\citeauthoryear{Coyne and Sproat}{Coyne and Sproat}{2022}]%
        {wordseye}
\bibfield{author}{\bibinfo{person}{Bob Coyne} {and} \bibinfo{person}{Richard
  Sproat}.} \bibinfo{year}{2022}\natexlab{}.
\newblock \bibinfo{title}{WordsEye: an automatic text-to-scene conversion
  system}.
\newblock
\newblock
\urldef\tempurl%
\url{https://doi.org/10.1145/383259.383316}
\showDOI{\tempurl}


\bibitem[\protect\citeauthoryear{Crowson}{Crowson}{2021a}]%
        {crowsondiffusion}
\bibfield{author}{\bibinfo{person}{Katherine Crowson}.}
  \bibinfo{year}{2021}\natexlab{a}.
\newblock \bibinfo{title}{afiaka87/clip-guided-diffusion: A CLI tool/python
  module for generating images from text using guided diffusion and CLIP from
  OpenAI.}
\newblock
\newblock
\urldef\tempurl%
\url{https://github.com/afiaka87/clip-guided-diffusion}
\showURL{%
\tempurl}


\bibitem[\protect\citeauthoryear{Crowson}{Crowson}{2021b}]%
        {crowson}
\bibfield{author}{\bibinfo{person}{Katherine Crowson}.}
  \bibinfo{year}{2021}\natexlab{b}.
\newblock \bibinfo{title}{Rivers Have Wings}.
\newblock
\newblock
\urldef\tempurl%
\url{https://twitter.com/RiversHaveWings}
\showURL{%
\tempurl}


\bibitem[\protect\citeauthoryear{Crowson, Biderman, Kornis, Stander, Hallahan,
  Castricato, and Raff}{Crowson et~al\mbox{.}}{2022}]%
        {crowson2022vqgan}
\bibfield{author}{\bibinfo{person}{Katherine Crowson}, \bibinfo{person}{Stella
  Biderman}, \bibinfo{person}{Daniel Kornis}, \bibinfo{person}{Dashiell
  Stander}, \bibinfo{person}{Eric Hallahan}, \bibinfo{person}{Louis
  Castricato}, {and} \bibinfo{person}{Edward Raff}.}
  \bibinfo{year}{2022}\natexlab{}.
\newblock \showarticletitle{VQGAN-CLIP: Open Domain Image Generation and
  Editing with Natural Language Guidance}.
\newblock \bibinfo{journal}{\emph{arXiv preprint arXiv:2204.08583}}
  (\bibinfo{year}{2022}).
\newblock


\bibitem[\protect\citeauthoryear{Cupchik, Vartanian, Crawley, and
  Mikulis}{Cupchik et~al\mbox{.}}{2009}]%
        {cupchik}
\bibfield{author}{\bibinfo{person}{Gerald Cupchik}, \bibinfo{person}{Oshin
  Vartanian}, \bibinfo{person}{Adrian Crawley}, {and} \bibinfo{person}{David
  Mikulis}.} \bibinfo{year}{2009}\natexlab{}.
\newblock \showarticletitle{Viewing artworks: Contributions of cognitive
  control and perceptual facilitation to aesthetic experience}.
\newblock \bibinfo{journal}{\emph{Brain and Cognition}}  \bibinfo{volume}{70}
  (\bibinfo{date}{06} \bibinfo{year}{2009}), \bibinfo{pages}{84--91}.
\newblock
\urldef\tempurl%
\url{https://doi.org/10.1016/j.bandc.2009.01.003}
\showDOI{\tempurl}


\bibitem[\protect\citeauthoryear{Dayma, Patil, Cuenca, Saifullah, Abraham,
  Le~Khac, Melas, and Ghosh}{Dayma et~al\mbox{.}}{2021}]%
        {dalle}
\bibfield{author}{\bibinfo{person}{Boris Dayma}, \bibinfo{person}{Suraj Patil},
  \bibinfo{person}{Pedro Cuenca}, \bibinfo{person}{Khalid Saifullah},
  \bibinfo{person}{Tanishq Abraham}, \bibinfo{person}{Phúc Le~Khac},
  \bibinfo{person}{Luke Melas}, {and} \bibinfo{person}{Ritobrata Ghosh}.}
  \bibinfo{year}{2021}\natexlab{}.
\newblock \bibinfo{title}{DALLE Mini}.
\newblock
\newblock
\urldef\tempurl%
\url{https://doi.org/10.5281/zenodo.1234}
\showDOI{\tempurl}


\bibitem[\protect\citeauthoryear{Devlin, Chang, Lee, and Toutanova}{Devlin
  et~al\mbox{.}}{2019}]%
        {devlin2019bert}
\bibfield{author}{\bibinfo{person}{Jacob Devlin}, \bibinfo{person}{Ming-Wei
  Chang}, \bibinfo{person}{Kenton Lee}, {and} \bibinfo{person}{Kristina
  Toutanova}.} \bibinfo{year}{2019}\natexlab{}.
\newblock \bibinfo{title}{BERT: Pre-training of Deep Bidirectional Transformers
  for Language Understanding}.
\newblock
\newblock
\showeprint[arxiv]{1810.04805}~[cs.CL]


\bibitem[\protect\citeauthoryear{El-Nouby, Sharma, Schulz, Hjelm, El~Asri,
  Kahou, Bengio, and Taylor}{El-Nouby et~al\mbox{.}}{2018}]%
        {keepdrawingit}
\bibfield{author}{\bibinfo{person}{Alaa El-Nouby}, \bibinfo{person}{Shikhar
  Sharma}, \bibinfo{person}{Hannes Schulz}, \bibinfo{person}{R~Devon Hjelm},
  \bibinfo{person}{Layla El~Asri}, \bibinfo{person}{Samira~Ebrahimi Kahou},
  \bibinfo{person}{Y. Bengio}, {and} \bibinfo{person}{Graham Taylor}.}
  \bibinfo{year}{2018}\natexlab{}.
\newblock \bibinfo{title}{Keep Drawing It: Iterative language-based image
  generation and editing}.
\newblock
\newblock


\bibitem[\protect\citeauthoryear{Esser, Rombach, and Ommer}{Esser
  et~al\mbox{.}}{[n.d.]}]%
        {vqgan}
\bibfield{author}{\bibinfo{person}{Patrick Esser}, \bibinfo{person}{Robin
  Rombach}, {and} \bibinfo{person}{Björn Ommer}.}
  \bibinfo{year}{[n.d.]}\natexlab{}.
\newblock \bibinfo{title}{Taming Transformers for High-Resolution Image
  Synthesis}.
\newblock
\newblock
\urldef\tempurl%
\url{https://doi.org/10.48550/ARXIV.2012.09841}
\showDOI{\tempurl}


\bibitem[\protect\citeauthoryear{Gao, Fisch, and Chen}{Gao
  et~al\mbox{.}}{2021}]%
        {gao_making_2021}
\bibfield{author}{\bibinfo{person}{Tianyu Gao}, \bibinfo{person}{Adam Fisch},
  {and} \bibinfo{person}{Danqi Chen}.} \bibinfo{year}{2021}\natexlab{}.
\newblock \showarticletitle{Making {Pre}-trained {Language} {Models} {Better}
  {Few}-shot {Learners}}.
\newblock \bibinfo{journal}{\emph{arXiv:2012.15723 [cs]}} (\bibinfo{date}{June}
  \bibinfo{year}{2021}).
\newblock
\urldef\tempurl%
\url{http://arxiv.org/abs/2012.15723}
\showURL{%
\tempurl}
\newblock
\shownote{arXiv: 2012.15723.}


\bibitem[\protect\citeauthoryear{Gatys, Ecker, and Bethge}{Gatys
  et~al\mbox{.}}{2015}]%
        {gatys2015neural}
\bibfield{author}{\bibinfo{person}{Leon~A. Gatys},
  \bibinfo{person}{Alexander~S. Ecker}, {and} \bibinfo{person}{Matthias
  Bethge}.} \bibinfo{year}{2015}\natexlab{}.
\newblock \bibinfo{title}{A Neural Algorithm of Artistic Style}.
\newblock
\newblock
\showeprint[arxiv]{1508.06576}~[cs.CV]


\bibitem[\protect\citeauthoryear{Ge and Parikh}{Ge and Parikh}{2021}]%
        {ge2021visual}
\bibfield{author}{\bibinfo{person}{Songwei Ge} {and} \bibinfo{person}{Devi
  Parikh}.} \bibinfo{year}{2021}\natexlab{}.
\newblock \bibinfo{title}{Visual Conceptual Blending with Large-scale Language
  and Vision Models}.
\newblock
\newblock
\showeprint[arxiv]{2106.14127}~[cs.CL]


\bibitem[\protect\citeauthoryear{Gero, Ashktorab, Dugan, Pan, Johnson, Geyer,
  Ruiz, Miller, Millen, Campbell, Kumaravel, and Zhang}{Gero
  et~al\mbox{.}}{2020}]%
        {gero2020}
\bibfield{author}{\bibinfo{person}{Katy~Ilonka Gero}, \bibinfo{person}{Zahra
  Ashktorab}, \bibinfo{person}{Casey Dugan}, \bibinfo{person}{Qian Pan},
  \bibinfo{person}{James Johnson}, \bibinfo{person}{Werner Geyer},
  \bibinfo{person}{Maria Ruiz}, \bibinfo{person}{Sarah Miller},
  \bibinfo{person}{David~R. Millen}, \bibinfo{person}{Murray Campbell},
  \bibinfo{person}{Sadhana Kumaravel}, {and} \bibinfo{person}{Wei Zhang}.}
  \bibinfo{year}{2020}\natexlab{}.
\newblock \bibinfo{booktitle}{\emph{Mental Models of AI Agents in a Cooperative
  Game Setting}}.
\newblock \bibinfo{publisher}{Association for Computing Machinery},
  \bibinfo{address}{New York, NY, USA}, \bibinfo{pages}{1–12}.
\newblock
\showISBNx{9781450367080}
\urldef\tempurl%
\url{https://doi.org/10.1145/3313831.3376316}
\showURL{%
\tempurl}


\bibitem[\protect\citeauthoryear{Gero and Chilton}{Gero and Chilton}{2019}]%
        {gero_metaphoria_2019}
\bibfield{author}{\bibinfo{person}{Katy~Ilonka Gero} {and}
  \bibinfo{person}{Lydia~B. Chilton}.} \bibinfo{year}{2019}\natexlab{}.
\newblock \showarticletitle{Metaphoria: {An} {Algorithmic} {Companion} for
  {Metaphor} {Creation}}. In \bibinfo{booktitle}{\emph{Proceedings of the 2019
  {CHI} {Conference} on {Human} {Factors} in {Computing} {Systems}}}.
  \bibinfo{publisher}{ACM}, \bibinfo{address}{Glasgow Scotland Uk},
  \bibinfo{pages}{1--12}.
\newblock
\showISBNx{978-1-4503-5970-2}
\urldef\tempurl%
\url{https://doi.org/10.1145/3290605.3300526}
\showDOI{\tempurl}


\bibitem[\protect\citeauthoryear{Gero, Liu, and Chilton}{Gero
  et~al\mbox{.}}{2021}]%
        {sparks}
\bibfield{author}{\bibinfo{person}{Katy~Ilonka Gero}, \bibinfo{person}{Vivian
  Liu}, {and} \bibinfo{person}{Lydia~B. Chilton}.}
  \bibinfo{year}{2021}\natexlab{}.
\newblock \bibinfo{title}{Sparks: Inspiration for Science Writing using
  Language Models}.
\newblock
\newblock
\urldef\tempurl%
\url{https://doi.org/10.48550/ARXIV.2110.07640}
\showDOI{\tempurl}


\bibitem[\protect\citeauthoryear{Jiang, Olson, Toh, Molina, Donsbach, Terry,
  and Cai}{Jiang et~al\mbox{.}}{2022}]%
        {promptbasedellen}
\bibfield{author}{\bibinfo{person}{Ellen Jiang}, \bibinfo{person}{Kristen
  Olson}, \bibinfo{person}{Edwin Toh}, \bibinfo{person}{Alejandra Molina},
  \bibinfo{person}{Aaron Donsbach}, \bibinfo{person}{Michael Terry}, {and}
  \bibinfo{person}{Carrie~J Cai}.} \bibinfo{year}{2022}\natexlab{}.
\newblock \bibinfo{title}{PromptMaker: Prompt-based Prototyping with Large
  Language Models}.
\newblock
\newblock
\urldef\tempurl%
\url{https://doi.org/10.1145/3491101.3503564}
\showDOI{\tempurl}


\bibitem[\protect\citeauthoryear{Karras, Laine, Aittala, Hellsten, Lehtinen,
  and Aila}{Karras et~al\mbox{.}}{2020}]%
        {karras2020analyzing}
\bibfield{author}{\bibinfo{person}{Tero Karras}, \bibinfo{person}{Samuli
  Laine}, \bibinfo{person}{Miika Aittala}, \bibinfo{person}{Janne Hellsten},
  \bibinfo{person}{Jaakko Lehtinen}, {and} \bibinfo{person}{Timo Aila}.}
  \bibinfo{year}{2020}\natexlab{}.
\newblock \bibinfo{title}{Analyzing and Improving the Image Quality of
  StyleGAN}.
\newblock
\newblock
\showeprint[arxiv]{1912.04958}~[cs.CV]


\bibitem[\protect\citeauthoryear{Kawabata and Zeki}{Kawabata and Zeki}{2004}]%
        {Kawabata2004NeuralCO}
\bibfield{author}{\bibinfo{person}{Hideaki Kawabata} {and}
  \bibinfo{person}{Semir Zeki}.} \bibinfo{year}{2004}\natexlab{}.
\newblock \showarticletitle{Neural correlates of beauty.}
\newblock \bibinfo{journal}{\emph{Journal of neurophysiology}}
  \bibinfo{volume}{91 4} (\bibinfo{year}{2004}), \bibinfo{pages}{1699--705}.
\newblock


\bibitem[\protect\citeauthoryear{Li and Liang}{Li and Liang}{2021}]%
        {li_prefix-tuning_2021}
\bibfield{author}{\bibinfo{person}{Xiang~Lisa Li} {and} \bibinfo{person}{Percy
  Liang}.} \bibinfo{year}{2021}\natexlab{}.
\newblock \showarticletitle{Prefix-{Tuning}: {Optimizing} {Continuous}
  {Prompts} for {Generation}}.
\newblock \bibinfo{journal}{\emph{arXiv:2101.00190 [cs]}} (\bibinfo{date}{Jan.}
  \bibinfo{year}{2021}).
\newblock
\urldef\tempurl%
\url{http://arxiv.org/abs/2101.00190}
\showURL{%
\tempurl}
\newblock
\shownote{arXiv: 2101.00190.}


\bibitem[\protect\citeauthoryear{Liu and Chilton}{Liu and Chilton}{[n.d.]}]%
        {liu_chilton}
\bibfield{author}{\bibinfo{person}{Vivian Liu} {and} \bibinfo{person}{Lydia
  Chilton}.} \bibinfo{year}{[n.d.]}\natexlab{}.
\newblock \bibinfo{title}{Neurosymbolic generation of 3D animal shapes through
  ... - ceur-ws.org}.
\newblock
\newblock
\urldef\tempurl%
\url{http://ceur-ws.org/Vol-2903/IUI21WS-HAIGEN-8.pdf}
\showURL{%
\tempurl}


\bibitem[\protect\citeauthoryear{Liu and Chilton}{Liu and Chilton}{2021}]%
        {liu2021design}
\bibfield{author}{\bibinfo{person}{Vivian Liu} {and} \bibinfo{person}{Lydia~B.
  Chilton}.} \bibinfo{year}{2021}\natexlab{}.
\newblock \bibinfo{title}{Design Guidelines for Prompt Engineering
  Text-to-Image Generative Models}.
\newblock
\newblock
\showeprint[arxiv]{2109.06977}~[cs.HC]


\bibitem[\protect\citeauthoryear{Matejka, Glueck, Bradner, Hashemi, Grossman,
  and Fitzmaurice}{Matejka et~al\mbox{.}}{2018}]%
        {Matejka2018}
\bibfield{author}{\bibinfo{person}{Justin Matejka}, \bibinfo{person}{Michael
  Glueck}, \bibinfo{person}{Erin Bradner}, \bibinfo{person}{Ali Hashemi},
  \bibinfo{person}{Tovi Grossman}, {and} \bibinfo{person}{George Fitzmaurice}.}
  \bibinfo{year}{2018}\natexlab{}.
\newblock \showarticletitle{Dream Lens: Exploration and Visualization of
  Large-Scale Generative Design Datasets}. In
  \bibinfo{booktitle}{\emph{Proceedings of the 2018 CHI Conference on Human
  Factors in Computing Systems}}. \bibinfo{publisher}{Association for Computing
  Machinery}, \bibinfo{address}{New York, NY, USA}, \bibinfo{pages}{1–12}.
\newblock
\showISBNx{9781450356206}
\urldef\tempurl%
\url{https://doi.org/10.1145/3173574.3173943}
\showDOI{\tempurl}


\bibitem[\protect\citeauthoryear{Murdock}{Murdock}{[n.d.]}]%
        {murdock2}
\bibfield{author}{\bibinfo{person}{Ryan Murdock}.}
  \bibinfo{year}{[n.d.]}\natexlab{}.
\newblock \bibinfo{title}{lucidrains/big-sleep: A simple command line tool for
  text to image generation, using OpenAI's CLIP and a BigGAN. Technique was
  originally created by https://twitter.com/advadnoun}.
\newblock
\newblock
\urldef\tempurl%
\url{https://github.com/lucidrains/big-sleep}
\showURL{%
\tempurl}


\bibitem[\protect\citeauthoryear{nerdyroden}{nerdyroden}{2022}]%
        {nerdyroden}
\bibfield{author}{\bibinfo{person}{nerdyroden}.}
  \bibinfo{year}{2022}\natexlab{}.
\newblock \bibinfo{title}{nerdyrodent/VQGAN-CLIP}.
\newblock
\newblock
\urldef\tempurl%
\url{https://github.com/nerdyrodent/VQGAN-CLIP}
\showURL{%
\tempurl}


\bibitem[\protect\citeauthoryear{Park, Liu, Wang, and Zhu}{Park
  et~al\mbox{.}}{2019}]%
        {park2019SPADE}
\bibfield{author}{\bibinfo{person}{Taesung Park}, \bibinfo{person}{Ming-Yu
  Liu}, \bibinfo{person}{Ting-Chun Wang}, {and} \bibinfo{person}{Jun-Yan Zhu}.}
  \bibinfo{year}{2019}\natexlab{}.
\newblock \showarticletitle{Semantic Image Synthesis with Spatially-Adaptive
  Normalization}. In \bibinfo{booktitle}{\emph{Proceedings of the IEEE
  Conference on Computer Vision and Pattern Recognition}}.
\newblock


\bibitem[\protect\citeauthoryear{Qiao, Zhang, Xu, and Tao}{Qiao
  et~al\mbox{.}}{2019a}]%
        {leicagan}
\bibfield{author}{\bibinfo{person}{Tingting Qiao}, \bibinfo{person}{Jing
  Zhang}, \bibinfo{person}{Duanqing Xu}, {and} \bibinfo{person}{Dacheng Tao}.}
  \bibinfo{year}{2019}\natexlab{a}.
\newblock \showarticletitle{Learn, Imagine and Create: Text-to-Image Generation
  from Prior Knowledge}. In \bibinfo{booktitle}{\emph{Advances in Neural
  Information Processing Systems}},
  \bibfield{editor}{\bibinfo{person}{H.~Wallach},
  \bibinfo{person}{H.~Larochelle}, \bibinfo{person}{A.~Beygelzimer},
  \bibinfo{person}{F.~d\textquotesingle Alch\'{e}-Buc},
  \bibinfo{person}{E.~Fox}, {and} \bibinfo{person}{R.~Garnett}} (Eds.),
  Vol.~\bibinfo{volume}{32}. \bibinfo{publisher}{Curran Associates, Inc.}
\newblock
\urldef\tempurl%
\url{https://proceedings.neurips.cc/paper/2019/file/d18f655c3fce66ca401d5f38b48c89af-Paper.pdf}
\showURL{%
\tempurl}


\bibitem[\protect\citeauthoryear{Qiao, Zhang, Xu, and Tao}{Qiao
  et~al\mbox{.}}{2019b}]%
        {mirrorgan}
\bibfield{author}{\bibinfo{person}{Tingting Qiao}, \bibinfo{person}{Jing
  Zhang}, \bibinfo{person}{Duanqing Xu}, {and} \bibinfo{person}{Dacheng Tao}.}
  \bibinfo{year}{2019}\natexlab{b}.
\newblock \bibinfo{title}{MirrorGAN: Learning Text-to-image Generation by
  Redescription}.
\newblock
\newblock
\urldef\tempurl%
\url{https://doi.org/10.48550/ARXIV.1903.05854}
\showDOI{\tempurl}


\bibitem[\protect\citeauthoryear{Radford, Kim, Hallacy, Ramesh, Goh, Agarwal,
  Sastry, Askell, Mishkin, Clark, Krueger, and Sutskever}{Radford
  et~al\mbox{.}}{2021}]%
        {radford2021learning}
\bibfield{author}{\bibinfo{person}{Alec Radford}, \bibinfo{person}{Jong~Wook
  Kim}, \bibinfo{person}{Chris Hallacy}, \bibinfo{person}{Aditya Ramesh},
  \bibinfo{person}{Gabriel Goh}, \bibinfo{person}{Sandhini Agarwal},
  \bibinfo{person}{Girish Sastry}, \bibinfo{person}{Amanda Askell},
  \bibinfo{person}{Pamela Mishkin}, \bibinfo{person}{Jack Clark},
  \bibinfo{person}{Gretchen Krueger}, {and} \bibinfo{person}{Ilya Sutskever}.}
  \bibinfo{year}{2021}\natexlab{}.
\newblock \bibinfo{title}{Learning Transferable Visual Models From Natural
  Language Supervision}.
\newblock
\newblock
\showeprint[arxiv]{2103.00020}~[cs.CV]


\bibitem[\protect\citeauthoryear{Ramesh, Dhariwal, Nichol, Chu, and
  Chen}{Ramesh et~al\mbox{.}}{2022}]%
        {unclip}
\bibfield{author}{\bibinfo{person}{Aditya Ramesh}, \bibinfo{person}{Prafulla
  Dhariwal}, \bibinfo{person}{Alex Nichol}, \bibinfo{person}{Casey Chu}, {and}
  \bibinfo{person}{Mark Chen}.} \bibinfo{year}{2022}\natexlab{}.
\newblock \bibinfo{title}{Hierarchical Text-Conditional Image Generation with
  CLIP Latents}.
\newblock
\newblock
\urldef\tempurl%
\url{https://doi.org/10.48550/ARXIV.2204.06125}
\showDOI{\tempurl}


\bibitem[\protect\citeauthoryear{{reddit.com}}{{reddit.com}}{2021}]%
        {R/bigsleep}
\bibfield{author}{\bibinfo{person}{{reddit.com}}.}
  \bibinfo{year}{2021}\natexlab{}.
\newblock
\newblock
\urldef\tempurl%
\url{https://www.reddit.com/r/bigsleep/}
\showURL{%
\tempurl}


\bibitem[\protect\citeauthoryear{Reimers and Gurevych}{Reimers and
  Gurevych}{2019}]%
        {sbert}
\bibfield{author}{\bibinfo{person}{Nils Reimers} {and} \bibinfo{person}{Iryna
  Gurevych}.} \bibinfo{year}{2019}\natexlab{}.
\newblock \bibinfo{title}{Sentence-BERT: Sentence Embeddings using Siamese
  BERT-Networks}.
\newblock
\newblock
\urldef\tempurl%
\url{https://doi.org/10.48550/ARXIV.1908.10084}
\showDOI{\tempurl}


\bibitem[\protect\citeauthoryear{Reynolds and McDonell}{Reynolds and
  McDonell}{2021}]%
        {reynolds2021prompt}
\bibfield{author}{\bibinfo{person}{Laria Reynolds} {and} \bibinfo{person}{Kyle
  McDonell}.} \bibinfo{year}{2021}\natexlab{}.
\newblock \bibinfo{title}{Prompt Programming for Large Language Models: Beyond
  the Few-Shot Paradigm}.
\newblock
\newblock
\showeprint[arxiv]{2102.07350}~[cs.CL]


\bibitem[\protect\citeauthoryear{Saharia, Chan, Saxena, Li, Whang, Denton,
  Ghasemipour, Ayan, Mahdavi, Lopes, Salimans, Ho, Fleet, and Norouzi}{Saharia
  et~al\mbox{.}}{2022}]%
        {imagen}
\bibfield{author}{\bibinfo{person}{Chitwan Saharia}, \bibinfo{person}{William
  Chan}, \bibinfo{person}{Saurabh Saxena}, \bibinfo{person}{Lala Li},
  \bibinfo{person}{Jay Whang}, \bibinfo{person}{Emily Denton},
  \bibinfo{person}{Seyed Kamyar~Seyed Ghasemipour},
  \bibinfo{person}{Burcu~Karagol Ayan}, \bibinfo{person}{S.~Sara Mahdavi},
  \bibinfo{person}{Rapha~Gontijo Lopes}, \bibinfo{person}{Tim Salimans},
  \bibinfo{person}{Jonathan Ho}, \bibinfo{person}{David~J Fleet}, {and}
  \bibinfo{person}{Mohammad Norouzi}.} \bibinfo{year}{2022}\natexlab{}.
\newblock \bibinfo{title}{Photorealistic Text-to-Image Diffusion Models with
  Deep Language Understanding}.
\newblock
\newblock
\urldef\tempurl%
\url{https://doi.org/10.48550/ARXIV.2205.11487}
\showDOI{\tempurl}


\bibitem[\protect\citeauthoryear{Sharma, Suhubdy, Michalski, Kahou, and
  Bengio}{Sharma et~al\mbox{.}}{2018}]%
        {chatpainter}
\bibfield{author}{\bibinfo{person}{Shikhar Sharma}, \bibinfo{person}{Dendi
  Suhubdy}, \bibinfo{person}{Vincent Michalski},
  \bibinfo{person}{Samira~Ebrahimi Kahou}, {and} \bibinfo{person}{Yoshua
  Bengio}.} \bibinfo{year}{2018}\natexlab{}.
\newblock \bibinfo{title}{ChatPainter: Improving Text to Image Generation using
  Dialogue}.
\newblock
\newblock
\urldef\tempurl%
\url{https://doi.org/10.48550/ARXIV.1802.08216}
\showDOI{\tempurl}


\bibitem[\protect\citeauthoryear{Shimizu, Fisher, Paris, McCann, and
  Fatahalian}{Shimizu et~al\mbox{.}}{2020}]%
        {shimizu}
\bibfield{author}{\bibinfo{person}{Evan Shimizu}, \bibinfo{person}{Matthew
  Fisher}, \bibinfo{person}{Sylvain Paris}, \bibinfo{person}{James McCann},
  {and} \bibinfo{person}{Kayvon Fatahalian}.} \bibinfo{year}{2020}\natexlab{}.
\newblock \showarticletitle{Design Adjectives: A Framework for Interactive
  Model-Guided Exploration of Parameterized Design Spaces}. In
  \bibinfo{booktitle}{\emph{Proceedings of the 33rd Annual ACM Symposium on
  User Interface Software and Technology}} (Virtual Event, USA)
  \emph{(\bibinfo{series}{UIST '20})}. \bibinfo{publisher}{Association for
  Computing Machinery}, \bibinfo{address}{New York, NY, USA},
  \bibinfo{pages}{261–278}.
\newblock
\showISBNx{9781450375146}
\urldef\tempurl%
\url{https://doi.org/10.1145/3379337.3415866}
\showDOI{\tempurl}


\bibitem[\protect\citeauthoryear{Singh, Bernal, Savchenko, and Glassman}{Singh
  et~al\mbox{.}}{2022}]%
        {stolenelephant}
\bibfield{author}{\bibinfo{person}{Nikhil Singh}, \bibinfo{person}{Guillermo
  Bernal}, \bibinfo{person}{Daria Savchenko}, {and} \bibinfo{person}{Elena~L.
  Glassman}.} \bibinfo{year}{2022}\natexlab{}.
\newblock \showarticletitle{Where to Hide a Stolen Elephant: Leaps in Creative
  Writing with Multimodal Machine Intelligence}.
\newblock \bibinfo{journal}{\emph{ACM Trans. Comput.-Hum. Interact.}}
  (\bibinfo{date}{jan} \bibinfo{year}{2022}).
\newblock
\showISSN{1073-0516}
\urldef\tempurl%
\url{https://doi.org/10.1145/3511599}
\showDOI{\tempurl}
\newblock
\shownote{Just Accepted.}


\bibitem[\protect\citeauthoryear{Wu, Jiang, Donsbach, Gray, Molina, Terry, and
  Cai}{Wu et~al\mbox{.}}{2022a}]%
        {promptchainer}
\bibfield{author}{\bibinfo{person}{Tongshuang Wu}, \bibinfo{person}{Ellen
  Jiang}, \bibinfo{person}{Aaron Donsbach}, \bibinfo{person}{Jeff Gray},
  \bibinfo{person}{Alejandra Molina}, \bibinfo{person}{Michael Terry}, {and}
  \bibinfo{person}{Carrie~J Cai}.} \bibinfo{year}{2022}\natexlab{a}.
\newblock \bibinfo{title}{PromptChainer: Chaining Large Language Model Prompts
  through Visual Programming}.
\newblock
\newblock
\urldef\tempurl%
\url{https://doi.org/10.48550/ARXIV.2203.06566}
\showDOI{\tempurl}


\bibitem[\protect\citeauthoryear{Wu, Terry, and Cai}{Wu et~al\mbox{.}}{2022b}]%
        {aichains}
\bibfield{author}{\bibinfo{person}{Tongshuang Wu}, \bibinfo{person}{Michael
  Terry}, {and} \bibinfo{person}{Carrie~J Cai}.}
  \bibinfo{year}{2022}\natexlab{b}.
\newblock \bibinfo{title}{AI Chains: Transparent and Controllable Human-AI
  Interaction by Chaining Large Language Model Prompts}.
\newblock
\newblock
\urldef\tempurl%
\url{https://doi.org/10.1145/3491102.3517582}
\showDOI{\tempurl}


\bibitem[\protect\citeauthoryear{Xia, Yang, Xue, and Wu}{Xia
  et~al\mbox{.}}{2020}]%
        {tedigan}
\bibfield{author}{\bibinfo{person}{Weihao Xia}, \bibinfo{person}{Yujiu Yang},
  \bibinfo{person}{Jing-Hao Xue}, {and} \bibinfo{person}{Baoyuan Wu}.}
  \bibinfo{year}{2020}\natexlab{}.
\newblock \bibinfo{title}{TediGAN: Text-Guided Diverse Face Image Generation
  and Manipulation}.
\newblock
\newblock
\urldef\tempurl%
\url{https://doi.org/10.48550/ARXIV.2012.03308}
\showDOI{\tempurl}


\bibitem[\protect\citeauthoryear{Xu, Zhang, Huang, Zhang, Gan, Huang, and
  He}{Xu et~al\mbox{.}}{2017}]%
        {Attngan}
\bibfield{author}{\bibinfo{person}{Tao Xu}, \bibinfo{person}{Pengchuan Zhang},
  \bibinfo{person}{Qiuyuan Huang}, \bibinfo{person}{Han Zhang},
  \bibinfo{person}{Zhe Gan}, \bibinfo{person}{Xiaolei Huang}, {and}
  \bibinfo{person}{Xiaodong He}.} \bibinfo{year}{2017}\natexlab{}.
\newblock \bibinfo{title}{AttnGAN: Fine-Grained Text to Image Generation with
  Attentional Generative Adversarial Networks}.
\newblock
\newblock
\urldef\tempurl%
\url{https://doi.org/10.48550/ARXIV.1711.10485}
\showDOI{\tempurl}


\bibitem[\protect\citeauthoryear{Yu, Xu, Koh, Luong, Baid, Wang, Vasudevan, Ku,
  Yang, Ayan, Hutchinson, Han, Parekh, Li, Zhang, Baldridge, and Wu}{Yu
  et~al\mbox{.}}{2022}]%
        {parti}
\bibfield{author}{\bibinfo{person}{Jiahui Yu}, \bibinfo{person}{Yuanzhong Xu},
  \bibinfo{person}{Jing~Yu Koh}, \bibinfo{person}{Thang Luong},
  \bibinfo{person}{Gunjan Baid}, \bibinfo{person}{Zirui Wang},
  \bibinfo{person}{Vijay Vasudevan}, \bibinfo{person}{Alexander Ku},
  \bibinfo{person}{Yinfei Yang}, \bibinfo{person}{Burcu~Karagol Ayan},
  \bibinfo{person}{Ben Hutchinson}, \bibinfo{person}{Wei Han},
  \bibinfo{person}{Zarana Parekh}, \bibinfo{person}{Xin Li},
  \bibinfo{person}{Han Zhang}, \bibinfo{person}{Jason Baldridge}, {and}
  \bibinfo{person}{Yonghui Wu}.} \bibinfo{year}{2022}\natexlab{}.
\newblock \bibinfo{title}{Scaling Autoregressive Models for Content-Rich
  Text-to-Image Generation}.
\newblock
\newblock
\urldef\tempurl%
\url{https://doi.org/10.48550/ARXIV.2206.10789}
\showDOI{\tempurl}


\end{thebibliography}

\appendix

\end{document}